\documentclass [12pt]{article}
\usepackage [cp1251]{inputenc}
\usepackage [english]{babel}
\usepackage {graphicx}
\usepackage {amssymb}
\usepackage {amsmath}
\usepackage {mathrsfs}
\usepackage {longtable}

\title{Weakly nonlinear magnetic convection in a nonuniformly  rotating electrically conductive medium under the action of modulation of external fields }
\author{$^{1}$\textbf{M.I. Kopp}, $^3$\textbf{A.V. Tur}, $^{1,2}$\textbf{V.V. Yanovsky}}

\begin{document}

\maketitle

$^{1}$ \textit{Institute for Single Crystals, NAS  Ukraine, Nauky Ave. 60, Kharkov 61001, Ukraine}

$^{2}$\textit{V.N. Karazin Kharkiv National University 4 Svobody Sq., Kharkov 61022, Ukraine}

$^{3}$\textit{Universit\'{e} de Toulouse [UPS], CNRS, Institut de Recherche en Astrophysique et Plan\'{e}tologie,
9 avenue du Colonel Roche, BP 44346, 31028 Toulouse Cedex 4, France}

\abstract{In this paper we studied the weakly nonlinear stage of stationary convective instability in a nonuniformly rotating layer of an electrically conductive fluid in an axial uniform magnetic field under the influence of: a) temperature modulation of the layer boundaries; b) gravitational modulation; c) modulation of the magnetic field; d) modulation of the angular velocity of rotation. As a result of applying the method of perturbation theory for the small parameter of supercriticality of the stationary Rayleigh number nonlinear non-autonomous Ginzburg-Landau equations for the above types of modulation were obtaned. By utilizing the solution of the Ginzburg-Landau equation, we determined the dynamics of unsteady heat transfer for various types of modulation of external fields and for different profiles of the angular velocity of the rotation of electrically conductive fluid.}

\section{Introduction}

As known, the instability of a horizontal fluid layer heated from below in the field of gravity (the Rayleigh-Benard convection) is a classic problem of fluid dynamics \cite{1s}-\cite{3s}. The problems related to the effect of rotation and magnetic field on the Rayleigh-Benard convection cause particular interest. These problems are of applied nature for astrophysical, geophysical and for engineering-technological research \cite{4s}. The problem of rotating Rayleigh-Benard convection was studied sufficiently detailed in \cite{5s}-\cite{6s}, where it was found that the Coriolis force with the rotation vector ${\bf{\Omega}}$ parallel to the gravity vector ${\bf{g}}$ inhibits the onset of convection and thus induces a stabilizing effect. Rayleigh-Benard convection, in which the axis of rotation of the medium and the uniform magnetic field coincide with the direction of the gravity vector, was well studied in \cite{1s}-\cite{2s}. The case is also interesting for astrophysical problems when the directions of the axes of rotation and the magnetic field are perpendicular to each other, and the direction of the magnetic field is perpendicular to the direction of the gravity vector. Such problem statement corresponds to convection in fluid layers located in the equatorial region of a rotating object, where the azimuthal magnetic field plays a significant role. The linear theory of such convection was first constructed in \cite{7s}-\cite{8s}. The linear theory of rotating magnetic convection for a random deviation of the axes of rotation and the magnetic field from the vertical axis (gravity field) was developed in \cite{9s}. A weakly nonlinear theory and stability analysis of azimuthal magnetic convection with $B_{0\phi}(R)=\textrm{const}$ was performed in \cite{10s}. It proposes a model in which the centrifugal acceleration $g_c=\Omega^2(R_1+R_2)$ can play the role of gravitational acceleration ${\bf{g}}$ for free convection in the local Cartesian approximation. The weakly nonlinear theory of centrifugal magnetoconvection considered in \cite{10s} was applied to the problem of a hydromagnetic dynamo. In all works on rotating magnetic convection \cite{1s}-\cite{10s}, the rotation of a horizontal fluid layer with a constant angular velocity ${\bf{\Omega}}=\textrm{const}$ was considered.

However, it is known that the majority of various space objects consisting of dense gases or liquid (Jupiter, Saturn, Sun, etc.) rotate non-uniformly, i.e. different parts of the object rotate around a common axis of rotation with different angular velocities. Differential (non-uniform) rotation is also observed in galaxies, accretion disks, and extended rings of planets. Besides, such large-scale vortex structures as typhoons, cyclones and anticyclones, etc. also rotate non-uniformly. This circumstance served as the motivation for a theoretical study of Rayleigh-Benard convection in a non-uniformly rotating electrically conductive fluid in the axial uniform magnetic field \cite{11s}-\cite{13s}, as well as in an external spiral magnetic field \cite{14s} with the nontrivial topology ${\bf{B}}_{0}\textrm{rot}{\bf{B}}_{0} \ne 0$.

The problem of the stability of an electrically conducting fluid between two rotating cylinders (Couette flow) and the Rayleigh-Benard problem in an external constant magnetic field were both considered in \cite{11s}-\cite{12s}. There was also carried out a study of the chaotic regime based on the equations of nonlinear dynamics of a six-dimensional $(6D)$ phase space. The analysis of these equations has shown the existence of a complex chaotic structure -- a strange attractor. A convection mode in which a chaotic change in direction (inversion) and amplitude of the perturbed magnetic field, taking into account the inhomogeneous rotation of the medium, occurs was found as well. A study of the chaotic regime of magnetic convection of a nonuniformly rotating electrically conductive fluid in a spiral magnetic field based on the equations of nonlinear dynamics of an eight-dimensional $(8D)$ phase space was carried out in \cite{14s}. There was also found a convection regime in which a chaotic change in direction (inversion) and amplitude of the perturbed magnetic field occurs, taking into account the nonuniform rotation of the medium and the nonuniform external azimuthal magnetic field. Earlier, a weakly nonlinear stage for rotating magnetoconvection (for $ \Omega=\textrm {const} $), in which a chaotic regime occurs, was studied in rotating fluid layers \cite{15s}-\cite{16s}, in conducting media with a uniform magnetic field \cite{17s}-\cite{20s}, and in conducting mediums rotating with a magnetic field \cite{21s}. However, the dynamics of the magnetic field itself was not considered in these works, which corresponds to the non-inductive approximation. Such tasks have great importance for technological applications: crystal growth, chemical processes of solidification and centrifugal casting of metals, etc.

The study of the dynamics of a magnetic field generated by convective motions of a fluid is important for the theory of magnetic dynamo \cite{22s}. A special role in this is played by issues related to the physical nature of inversions and variations in the magnetic field of the Earth, the Sun, and other space objects. In \cite{23s} Rikitaki proposed an electromechanical model of terrestrial magnetism. The study of the dynamic system of Rikitaki equations was also used to explain the chaotic inversion of the geomagnetic field \cite{24s}-\cite{27s}. In recent works \cite{28s}-\cite{29s} was investigated a modified system of Rikitaki equations taking into account friction and not reducing it to a three-dimensional form as for example in \cite{24s}. This made it possible to more clearly show that at first the oscillations of the current (or magnetic) variable near a certain stationary state with an increase in amplitude go into oscillations around an another stationary state, which simulated inversions \cite{29s}. In \cite{28s} it is established that after chaotic behavior the system goes into stable mode. According to the authors of \cite{28s}, such a regime can describe superchrons in the inversion of the geomagnetic field. In contrast to the works \cite{23s}-\cite{29s}, in \cite{11s}-\cite{12s}, \cite{14s} it is proposed to model the magnetic field inversion by a dynamic system of equations of Lorentz type, respectively, for $(6D)$ and $(8D)$- dimensional phase space.

In \cite{13s}, the weakly nonlinear stage of stationary convective instability in a nonuniformly rotating layer of an electrically conductive fluid in an axial uniform magnetic field was studied. As a result of applying the perturbation theory method for a small parameter of supercriticality of the stationary Rayleigh number \cite{30s}, the nonlinear autonomous Ginzburg-Landau equation was obtained. This equation describes the evolution of the finite amplitude of perturbations. A numerical analysis of this equation showed that the heat flux increases with rotation of the medium with positive Rossby numbers $\textrm{Ro}> 0$. In \cite{13s} it is shown that the weakly nonlinear convection based on the equations of the six-mode $(6D)$ Lorentz model transforms into the identical Ginzburg-Landau equation. The weakly nonlinear theory of convection was especially developed with regard to modulation of the parameters that control the convection process, what is very important for solving many technological problems. Different types of modulation, such as rotation \cite{31s}-\cite{34s}, gravity \cite{35s}-\cite{37s}, temperature \cite{38s}-\cite{40s} and magnetic field \cite{41s}-\cite{42s}, were studied for stationary weakly nonlinear convection in various media: porous media, nanofluids, and so on. In these papers \cite{31s}-\cite{42s} the effect of modulation of the parameters (rotation, gravity, temperature, magnetic field) on the heat and mass transfer in convective media was determined. A parametric effect on convection can lead to either an increase or a decrease in heat transfer. In addition to technological problems, considering of the modulation of external fields plays an important role in modeling convective processes on the Earth, the Sun, and other space objects \cite{43s}-\cite{44s}.

The content of the work is outlined in the following sections. The basic equations for the evolution of small perturbations in the Boussinesq approximation, that describe non-uniformly rotating convection in external periodic fields: a) temperature modulation of the layer boundaries, b) gravitational modulation, c) modulation of the magnetic field, d) modulation of the angular velocity of rotation, are obtained in Section 2. In Section 3 we study the weakly nonlinear stage of stationary convection in a nonuniformly rotating layer of an electrically conductive fluid under the action of modulation of external fields. Using the method of perturbation theory with respect to the small parameter of supercriticality of the Rayleigh number $ \epsilon = \sqrt {(\textrm {Ra}-\textrm {Ra}_c) / \textrm {Ra}_c} $) we obtained the nonlinear Ginzburg-Landau equation with a periodic coefficient for each type of modulation. The results of numerical solutions of the non-autonomous Ginzburg-Landau equation for each type of modulation show the dependence of the heat transfer (Nusselt number $ \textrm{Nu} $) on the amplitude $ \delta_{\textrm{mod}} $, the frequency  $ \omega_{\textrm{mod}} $ of the modulation and the profile of the nonuniformly rotation (number Rossby $ \textrm{Ro} $) are also presented in Section 3.

The results developed in this work can be applied to various astrophysical and geophysical problems that consider magnetic convection in the rotating layers of the Sun, hot galactic clusters, accretion disks and other objects.

\begin{figure}
  \centering
\includegraphics[width=12 cm, height=7 cm]{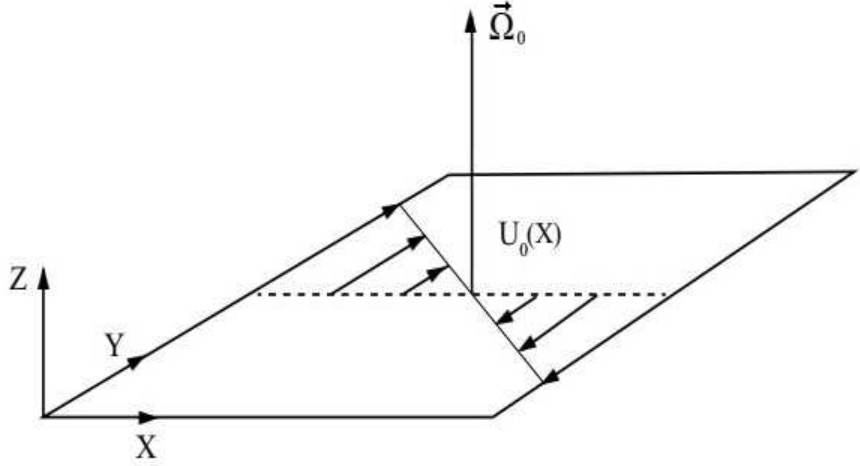}  \\
	\caption{Scheme of the shear flow in rotating flows, the flow being approximated in the local Cartesian coordinate system as a linear shift with velocity ${\bf{U}}_0(X)$.}\label{fg1}
\end{figure}

\section {Problem statement and basic evolution equations}

Let us consider a nonuniformly rotating flow of an electrically conductive fluid located between two impermeable horizontal planes $z=0$ and $z=h$, which are heated from below and cooled from above according to the periodic law. The temperature of the lower and upper horizontal boundaries is modulated in accordance with a time-harmonic law: 
\begin{equation} \label{eq1n} 
T_1=T_0+\frac{\Delta T}{2}[1+\epsilon^2\delta_1\cos(\widetilde{\omega}_T t)] \quad \textrm{at} \quad z=0
\end{equation}
\[ T_2=T_0-\frac{\Delta T}{2}[1-\epsilon^2\delta_1\cos(\widetilde{\omega}_T t+\varphi)] \quad \textrm{at} \quad z=h, \]
where $ T_0 = \textrm {const} $  is the temperature relative to which oscillations occur with a frequency of $ \widetilde {\omega}_T $ and a phase shift of $ \varphi $, $ \Delta T $ is the temperature difference between the lower and upper planes in the absence of modulation, $ \delta_1 $ is the amplitude of thermal modulation, $ \epsilon$ is a small parameter. In a cylindrical coordinate system an electrically conductive medium (plasma) rotates in the azimuthal direction with the speed $v_{\phi} = R \Omega (R, t) $.  Here  $ \Omega (R, t) $ is the angular velocity of rotation, which makes small oscillations in time according to the periodic law: 
\begin{equation} \label{eq2n} 
  \Omega (R,t)=\Omega (R)(1+\epsilon^2\delta_2\cos(\widetilde{\omega}_R t)),
\end{equation}
where $ \widetilde {\omega}_R $ is the frequency of rotation modulation, $ \delta_2 $ is the amplitude of rotational modulation.

It is convenient to switch from a cylindrical coordinate system $(R,\varphi,z)$ to a local Cartesian system $(X,Y,Z)$ in order to describe nonlinear convective phenomena in a nonuniformly rotating layer of an electrically conducting fluid. If we consider a fixed region of a fluid layer with a radius $R_0$  and an angular velocity of rotation 
$$\Omega_0(t)=\Omega(R_0,t)=\Omega_{00}(1+\epsilon^2\delta_2\cos(\widetilde{\omega}_R t)), \quad \Omega_{00}=\textrm{const}, $$
then the coordinates $X=R-R_0$  correspond to the radial direction, $Y=R_0(\varphi-\varphi_0)$ to the azimuth and $Z=z$ -- to the vertical direction (see Fig. \ref{fg1}).
\begin{figure}
  \centering
\includegraphics[width=17 cm, height=16 cm]{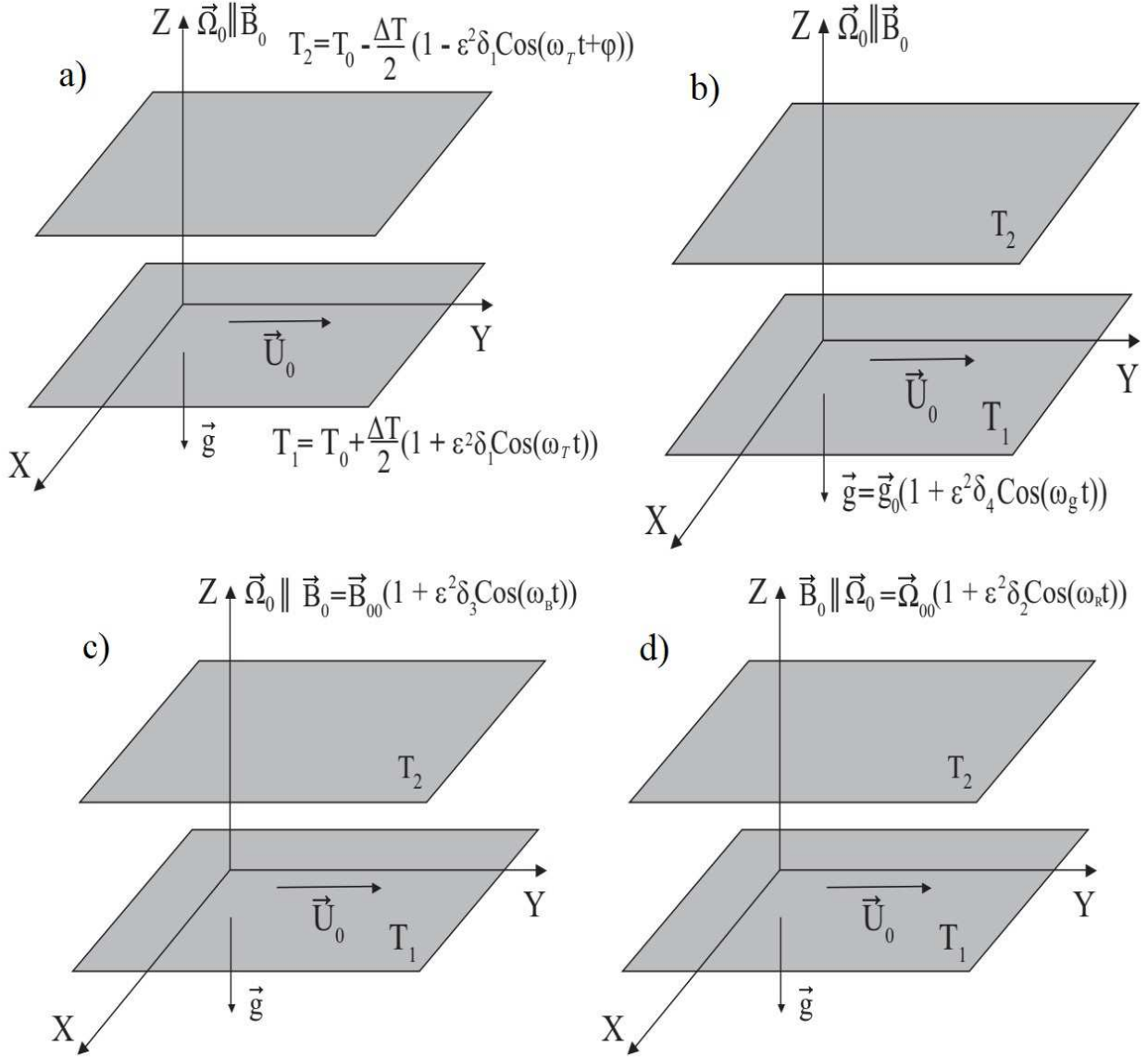}  \\
	\caption{ Cartesian approximation for a nonuniformly rotating magnetoconvection under parametric influence: $ \textrm {a)} $ temperature modulation of the boundaries of the liquid layer; $ \textrm {b)} $ modulation of the gravity field; $ \textrm {c)} $ modulation of the external magnetic field; $ \textrm {d)} $ modulation of the angular velocity of rotation. A nonuniformly rotation in the local Cartesian coordinate system consists of rotation with a constant angular velocity $ {\bf {\Omega}}_0 $ and shear velocity $ {\bf {U}}_0 \| \textrm{OY} $.}\label{fg2}
\end{figure}

Then, nonuniform  rotation of the fluid layer can be represented locally as a rotation with a constant angular velocity  ${\bf{\Omega}}_0(t)$   and azimuthal width \cite{46s}, which velocity profile is locally linear: ${\bf{U}}_0=-q {\Omega}_0(t) X {\bf{e}}_y$, where $q\equiv -d \ln \Omega/d\ln R$  is the dimensionless  shear flow parameter defined using the profile of angular velocity of rotation $\Omega (R)=\Omega_0 (R/R_0)^{-q}$. The parameter $q$ is related to the hydrodynamic Rossby number $\textrm{Ro}=\frac{R}{2\Omega} \frac{\partial\Omega}{\partial R}$ by the relation: $q=-2\textrm{Ro}$. Note that the accretion disks with the shear flow parameter $q=3/2$ $(\textrm{Ro}=-3/4)$ correspond to the Keplerian disk, $q=2$ $(\textrm{Ro}=-1) $  corresponds to the disk with a constant angular momentum or Rayleigh rotation profile. The case $q=1$ $(\textrm{Ro}=-1/2)$  corresponds to a system with a flat rotation curve, while $q=0$ $(\textrm{Ro}=0)$  corresponds to a uniform (or solid-body) rotation with a constant angular velocity.

We assume that the direction of the external magnetic field $ {\bf B}_{0} $ coincides with the axis of rotation of the fluid $ {\bf \Omega} $ $ \parallel OZ $. In addition, the external magnetic field $ {\bf B}_{0} $ and the gravitational acceleration vector $ {\bf g} = (0,0, -g) $ change with time according to the harmonic law
\begin{equation} \label{eq3n} 
 {\bf B}_{0} = B_{00}(1+ \epsilon^2 \delta_3 \cos(\widetilde{\omega}_B t)){\bf e}_Z, $$
 $$ {\bf g}=-g_{0}(1+\epsilon^2 \delta_4 \cos (\widetilde{\omega}_g t)){\bf e}_Z,  
\end{equation}
where $ \delta_3, \delta_4 $ are small amplitudes of magnetic and gravitational modulation, $ B_{00} = \textrm {const} $, $ \widetilde {\omega}_B, \widetilde {\omega}_g $ are frequencies modulation of magnetic and gravitational fields.

The influence of modulation of external fields  is considered on the basis of the equations of magnetohydrodynamics in the Boussinesq approximation \cite{1s}-\cite{2s}:
\begin{equation} \label{eq4n} \frac{\partial {\bf{v}}}{\partial t} +({\bf{v}}\nabla ){\bf{v}}=-\frac{1}{\rho _{0} } \nabla (P+\frac{B^{2} }{8\pi } )+\frac{1}{4\pi \rho _{0} } ({\bf{B}}\nabla ){\bf{B}}+g\beta T{\bf{e}}_Z+\nu \nabla^2 {\bf{v}} \end{equation} 
\begin{equation} \label{eq5n} \frac{\partial {\bf{B}}}{\partial t} +({\bf{v}}\nabla ){\bf{B}}-({\bf{B}}\nabla ){\bf{v}}=\eta \nabla^2 {\bf{B}} \end{equation} 
\begin{equation} \label{eq6n} \frac{\partial T}{\partial t} +({\bf{v}}\nabla )T=\chi \nabla^2 T \end{equation} 
\begin{equation} \label{eq7n} \textrm{div}{\bf{B}}=0, \quad \textrm{div}{\bf{v}}=0, \end{equation}
where $ {\bf {e}}_Z $ is the unit vector directed vertically up the $ OZ $ axis, $ \beta $ is the coefficient of thermal expansion, $ \rho_{0}=\textrm {const} $ is the density of the medium, $ \nu $ is the kinematic viscosity coefficient, $ \eta = c^2/4 \pi \sigma $ is the magnetic viscosity coefficient, $ \sigma $ is the conductivity coefficient, $ \chi $ is the coefficient thermal conductivity of the medium.
Let us represent all quantities in Eqs. (\ref{eq4n})-(\ref{eq7n}) as the sum of the stationary and perturbed components ${\bf{v}}={\bf{U}}_0+{\bf{u}}$, ${\bf{B}}={\bf{B}}_0+{\bf{b}}$, $P=p_0+p$, $T=T_b+\theta$.
The equations for the stationary state are:
\begin{equation} \label{eq8n} \frac{d p_0}{d Z}=\rho_0 g\beta T_b
\end{equation}
\begin{equation} \label{eq9n} -2q\Omega_0^2 X= \frac{1}{\rho_0}\frac{d p_0}{d X} \end{equation}
\begin{equation} \label{eq10n} \frac{\partial T_b}{\partial t} = \frac{d^2 T_b}{d Z^2} \end{equation}
\begin{figure}
  \centering
	\includegraphics[width=12 cm, height=8 cm]{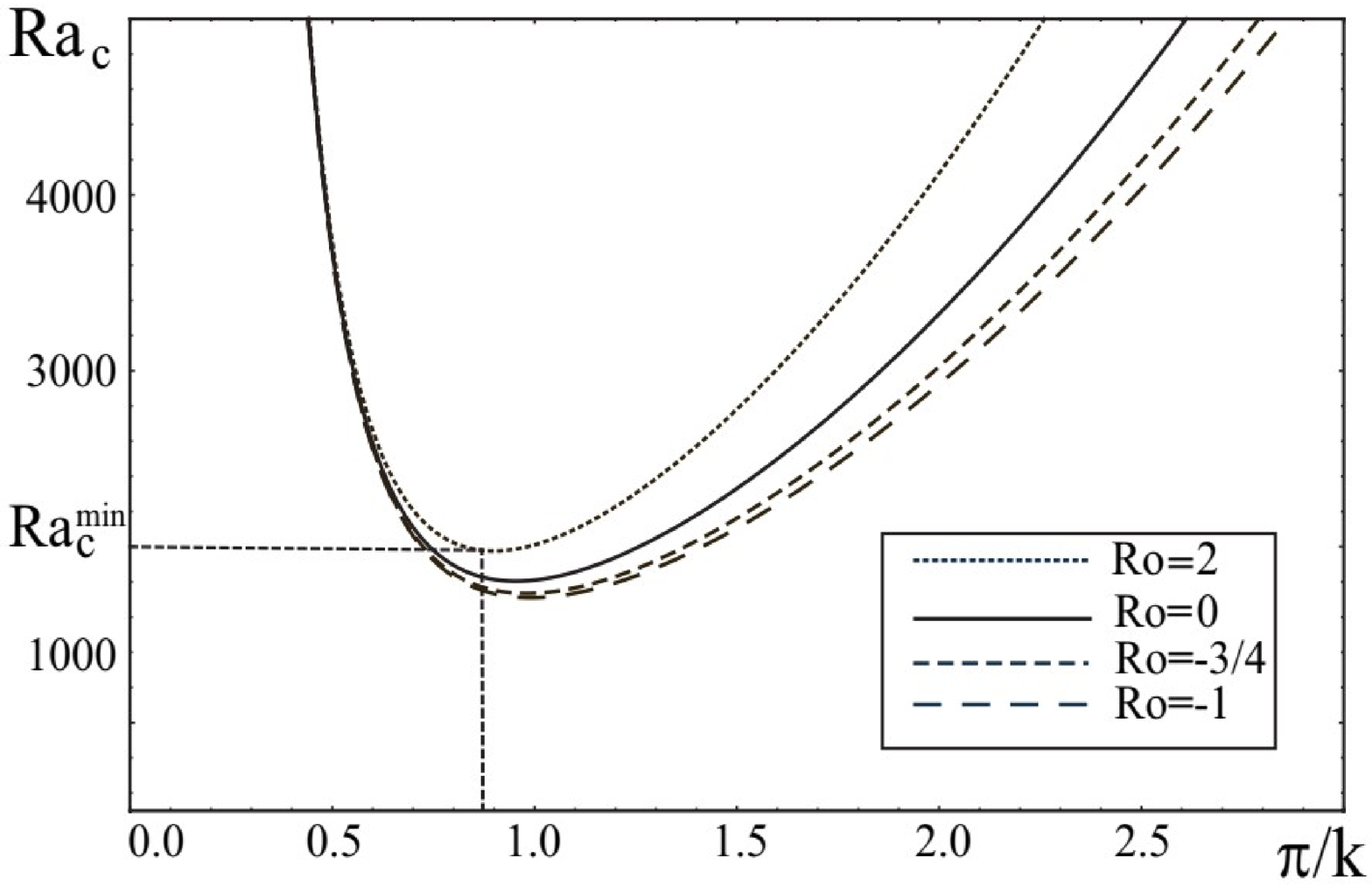}\\
\caption{ Dependences of  critical Rayleigh number $ \textrm{Ra}_c $ on  wavenumbers  $ \pi / k $ for different Rossby numbers $ \textrm {Ro} $ for constant parameters $ \textrm {Q} = 50 $, $ \textrm {Ta} = 100$  and $ \textrm {Pm} = 1 $.}\label{fg3}
\end{figure}
The expressions (\ref{eq8n})-(\ref{eq9n}) show that centrifugal equilibrium is established in the radial direction and hydrostatic in the vertical direction. The solution of the equation (\ref{eq10n}) with boundary conditions (\ref{eq1n}) has the form \cite{45s}:
\begin{equation} \label{eq11n} T_b(z,t)=T_S(z)+\epsilon^2\delta_1\cdot f_1(z,t)\Delta T,  \end{equation}
where
\[ T_S(z)=T_0+\frac{\Delta T}{2}\left(1-\frac{2z}{h}\right), \]
\[ f_1(z,t)=\textrm{Re} \left\{\left(a(\lambda)e^{\frac{\lambda z}{h}} +a(-\lambda)e^{-\frac{\lambda z}{h}}\right)e^{-i\widetilde{\omega}_T t}\right\}, \; \lambda^2=-\frac{i\widetilde{\omega}_T h^2}{\chi},\; a(\lambda)=\frac{1}{2}\cdot\frac{e^{-i\varphi}-e^{-\lambda}}{e^{\lambda}-e^{-\lambda}},\]
where $ T_S (z) $ is the stationary temperature, $ f_1 (z, t) $ is the oscillating part of $ T_b $, the symbol $ \textrm {Re} $ denotes the real part.

Subtracting the equations for the stationary state (\ref{eq8n})-(\ref{eq10n}) from (\ref{eq4n})-(\ref{eq7n}) we can find the evolution equations for small perturbations: 
\[\frac{{\partial {\bf u}}}{{\partial t}} -q {\Omega}_0 X\frac{\partial {\bf u}}{\partial Y}  + ({\bf u}\nabla ){\bf U}_0 +2{\bf{\Omega}}_0\times {\bf{u}}+ ({\bf u}\nabla ){\bf u}  =  - \frac{1}{{\rho _0 }}\nabla \widetilde p + \frac{1}{{4\pi \rho _0 }}\left(({\bf B}_0 \nabla ){\bf b} +({\bf b}\nabla ){\bf b}  \right) + g\beta \theta {\bf{e}}_Z + \nu \nabla^2 {\bf u}\]
\begin{equation} \label{eq12n} \frac{{\partial {\bf b}}}{{\partial t}} -q {\Omega}_0 X\frac{\partial {\bf b}}{\partial Y} - ({\bf B}_0 \nabla ){\bf u} - ({\bf b}\nabla ){\bf U}_0 +({\bf u}\nabla ){\bf b}-({\bf b}\nabla ){\bf u} = \eta \nabla^2 {\bf b} \end{equation}  
\[\frac{{\partial \theta }}{{\partial t}} -q {\Omega}_0 x\frac{\partial \theta}{\partial y}  + ({\bf u}\nabla )T_b+({\bf u}\nabla )\theta  = \chi \nabla^2 \theta \]
\[\textrm{div}{\bf{b}}=0, \quad \textrm{div}{\bf{u}}=0\]
\begin{figure}
  \centering
	\includegraphics[width=12 cm, height=8 cm]{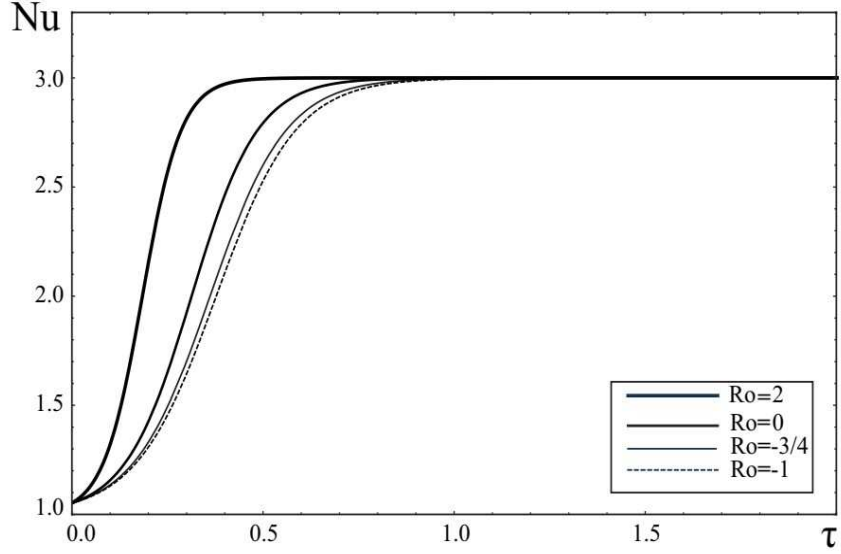} \\
\caption{ Dependency of the Nusselt number $ \textrm {Nu} $ on time $ \tau $ for Rossby numbers $ \textrm {Ro} = (2, 0, -3/4, -1) $ in the case of in-phase $ (\textrm {IPM} , \varphi = 0) $ of temperature modulation for a frequency $ \omega_T = 2 $  and an amplitude $ \delta_1 = 0.5 $. }\label{fg3t}
\end{figure}
Here the pressure $\widetilde{p}$ includes the disturbed magnetic pressure $ p_m=\frac{{\bf b}^2}{8\pi }+\frac{{\bf b}{ \bf B}_0}{4\pi}$:  $\widetilde{p}=p+p_m$. 

Let us consider the dynamics of axisymmetric perturbations, then all the perturbed quantities in the equations (\ref{eq11n}) will depend only on  two variables $ (X, Z) $:
\[{\bf{u}}=(u(X,Z) ,v(X,Z),w(X,Z)),\; {\bf{b}}=(\widetilde u(X,Z), \widetilde v(X,Z), \widetilde w(X,Z)),\; \widetilde p=\widetilde p(X,Z),\; \theta=\theta (X,Z)\]
The solenoidal equations for axisymmetric velocity and magnetic field perturbations will take the form
\begin{equation} \label{eq13n} \frac{\partial u}{\partial X}+\frac{\partial w}{\partial Z}=0, \quad \frac{\partial \widetilde u}{\partial X}+\frac{\partial \widetilde w}{\partial Z}=0
\end{equation}	
The remaining equations in the coordinate representation will take the following form:
\begin{equation} \label{eq14n} 
\left(\frac{\partial}{\partial t}-\nu\nabla^2\right)u+({\bf u} \nabla)u=-\frac{1}{\rho_0}\frac{\partial \widetilde p}{\partial X}+2\Omega_{00}f_R v+\frac{1}{4\pi \rho_0}({\bf b} \nabla)\widetilde u+\frac{B_{00}f_m}{4\pi \rho_0}\frac{\partial \widetilde u}{\partial Z}
\end{equation}
\begin{equation} \label{eq15n} 
\left(\frac{\partial}{\partial t}-\nu\nabla^2\right)v+({\bf u} \nabla)v=-2\Omega_{00}f_Ru\left(1-\frac{q}{2}\right)+\frac{1}{4\pi \rho_0}({\bf b} \nabla)\widetilde v+ \frac{B_{00}f_m}{4\pi \rho_0}\frac{ \partial \widetilde v }{\partial Z}
\end{equation}
\begin{equation} \label{eq16n} 
\left(\frac{\partial}{\partial t}-\nu\nabla^2\right)w+({\bf u} \nabla)w=-\frac{1}{\rho_0}\frac{\partial \widetilde p}{\partial Z}+g_0f_g\beta\theta+\frac{1}{4\pi \rho_0}({\bf b} \nabla)\widetilde w+\frac{B_{00}f_m}{4\pi\rho_0}\frac{\partial \widetilde w}{\partial Z}
\end{equation}
\begin{equation} \label{eq17n}
\left(\frac{\partial}{\partial t}-\eta\nabla^2\right)\widetilde u-B_{00}f_m\frac{\partial u}{\partial Z}+({\bf u} \nabla)\widetilde u-({\bf b} \nabla)u=0
\end{equation}
\begin{equation} \label{eq18n}
\left(\frac{\partial}{\partial t}-\eta\nabla^2\right)\widetilde v-B_{00}f_m\frac{\partial v}{\partial Z}+q\Omega_{00}f_R\widetilde u +({\bf u} \nabla)\widetilde v-({\bf b} \nabla)v=0
\end{equation}
\begin{figure}
  \centering
	\includegraphics[width=12 cm, height=8 cm]{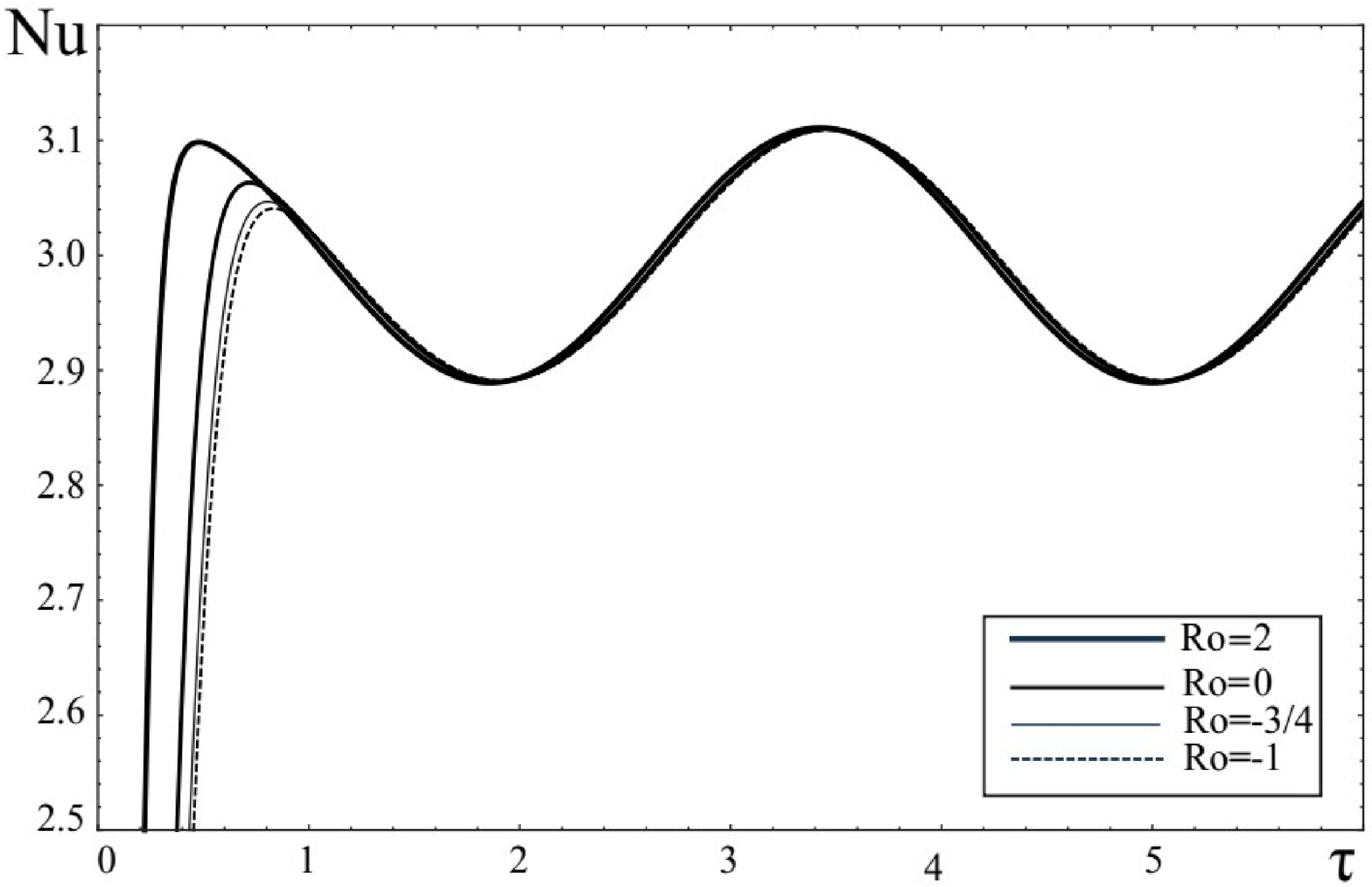}\\
\caption{Dependency of the Nusselt number $ \textrm{Nu} $ on the time $ \tau $ for Rossby numbers $ \textrm {Ro} = (2, 0, -3/4, -1) $ for the case of phase $ (\textrm{OPM} , \varphi = \pi) $ of temperature modulation for frequency of $ \omega_T = 2 $ and amplitude of $ \delta_1 = 0.5 $. }\label{fg3tf}
\end{figure}
\begin{equation} \label{eq19n}
\left(\frac{\partial}{\partial t}-\eta\nabla^2\right)\widetilde w-B_{00}f_m\frac{\partial w}{\partial Z}+({\bf u} \nabla)\widetilde w-({\bf b} \nabla)w=0
\end{equation}
\begin{equation} \label{eq20n}
\left(\frac{\partial}{\partial t}-\chi\nabla^2\right)\theta-Aw+\epsilon^2\delta_1\cdot\frac{\partial f_1}{\partial Z}\Delta T w+(\vec u \nabla)\theta=0, \; A=\frac{\Delta T}{h},
\end{equation}
where $f_R=1+\epsilon^2\delta_2\cos \widetilde{\omega}_R t,  f_m=1+\epsilon^2\delta_3\cos \widetilde{\omega}_B t, f_g=1+\epsilon^2\delta_4\cos \widetilde{\omega}_g t$.

In the equations (\ref{eq14n})-(\ref{eq20n}) the nabla operators can be described as:
$$({\bf a} \nabla)=a_x\,\frac{\partial}{\partial X}+a_z\,\frac{\partial}{\partial Z},\quad \nabla^2=\frac{\partial^2}{\partial X^2}+\frac{\partial^2}{\partial Z^2}. $$
To eliminate the pressure $ \widetilde{p} $ in the equations (\ref{eq14n}) and (\ref{eq16n}), we need to differentiate the equation (\ref{eq14n}) with respect to $Z $.  The equation (\ref{eq16n}) is required to be differentiated  with respect to $ X $  and then by subtracting them from each other, we can obtain the equation for $ Y $-the components of the vortex $ \textrm{rot} {\bf u} = {\bf e}_Y \omega $:
\begin{equation} \label{eq21n} 
\left(\frac{\partial}{\partial t}-\nu\nabla^2\right)\omega+\frac{\partial}{\partial Z}\left(u\frac{\partial u}{\partial X}+w\frac{\partial u}{\partial Z}\right)-\frac{\partial}{\partial X}\left(u\frac{\partial w}{\partial X}+w\frac{\partial w}{\partial Z}\right)= \frac{B_{00}f_m}{4\pi\rho_0}\frac{\partial I}{\partial Z}+$$
$$+\frac{1}{4\pi\rho_0}\left(\frac{\partial}{\partial Z}\left(\widetilde u \frac{\partial \widetilde u}{\partial X}+ \widetilde w\frac{\partial \widetilde u}{\partial Z}\right)-\frac{\partial}{\partial X}\left(\widetilde u \frac{\partial \widetilde w}{\partial X}+\widetilde w \frac{\partial \widetilde w}{\partial Z}\right)\right) + 2\Omega_{00}f_R\frac{\partial  v}{\partial Z}-g_0f_g\beta\frac{\partial\theta}{\partial X}
\end{equation}
where $ \omega = \frac {\partial u} {\partial Z}-\frac {\partial w} {\partial X} $ is the $ Y $ -component of the vortex, $ I = \frac {\partial \widetilde u} {\partial Z} - \frac {\partial \widetilde w} {\partial X} $ - $ Y $ -current component $ {\bf I} = \textrm {rot} {\bf b} = I {\bf e} _Y $.
According to the equations (\ref{eq13n}) it is convenient to introduce the stream function $ \psi $ through which the components of the perturbed velocity are expressed:
 \[ u=-\frac{\partial\psi}{\partial Z},\quad w=\frac{\partial\psi}{\partial X} \]
\begin{figure}
 \centering
\includegraphics[width=12 cm, height=8 cm]{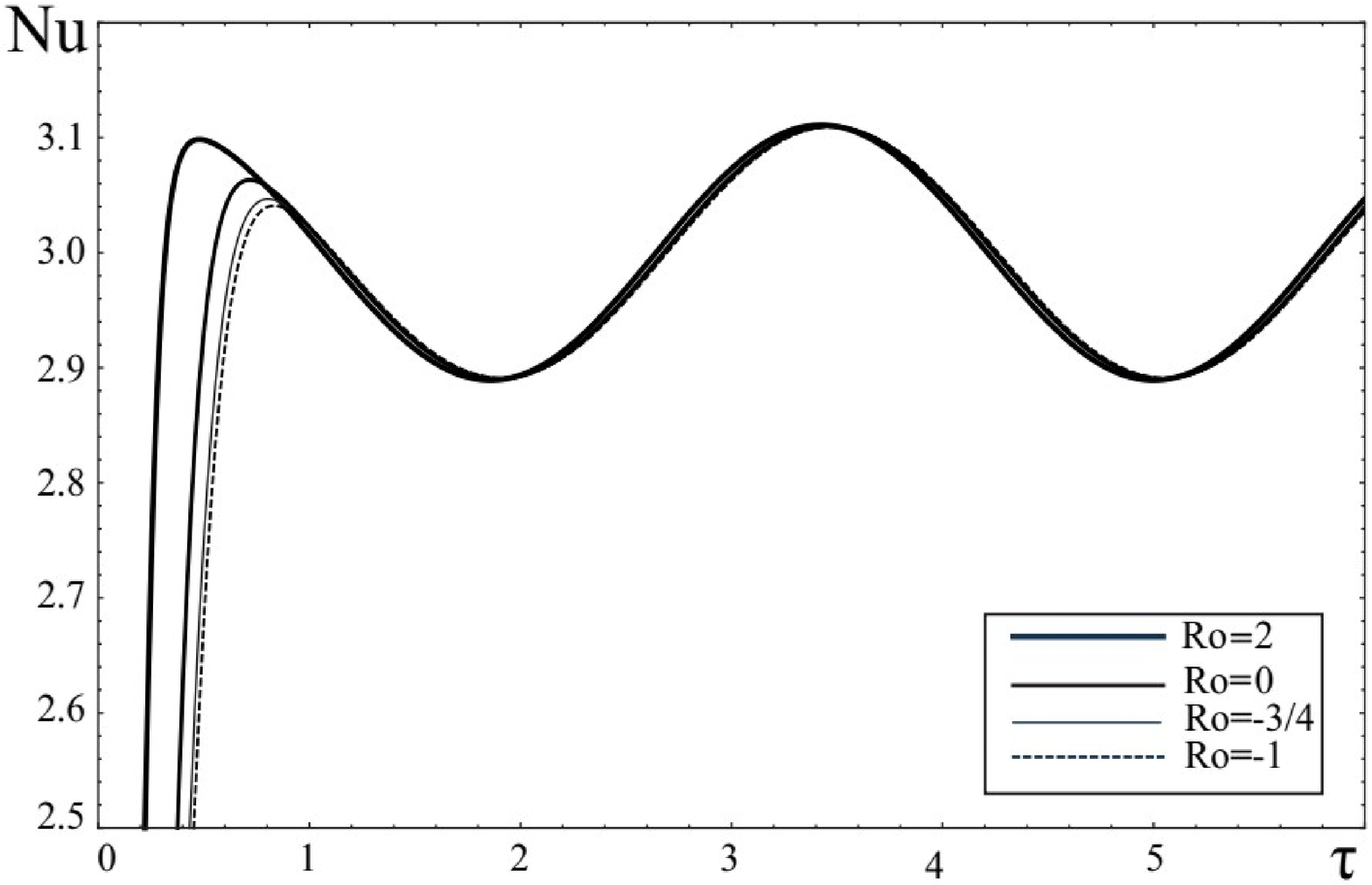}\\
\caption{Dependency of the Nusselt number $ \textrm {Nu} $ on the time $ \tau $ for Rossby numbers $ \textrm {Ro} = (2, 0, -3/4, -1) $ for the case of temperature modulation of only the lower boundary of the layer $ ( \textrm {LBMO}, \varphi = -i \infty) $ for frequency $ \omega_T = 2 $ and amplitude $ \delta_1 = 0.5 $.}\label{fg3tfm}
\end{figure}	
Similarly, we can introduce the stream function $ \phi $ for perturbations of the magnetic field:
 \[ \widetilde u=-\frac{\partial\phi}{\partial Z},\quad \widetilde w=\frac{\partial\phi}{\partial X} \]
As a result, the equations (\ref{eq21n}) and (\ref{eq15n}) become more compact
\begin{equation} \label{eq22n} 
 \left(\frac{\partial}{\partial t}-\nu\nabla^2\right) \nabla ^2 \psi  + 2\Omega_{00}f_R \frac{{\partial v}}{{\partial Z}} - \frac{{B_{00}f_m }}{{4\pi \rho _0 }}\frac{\partial }{{\partial Z}}\nabla ^2 \phi  - g_0f_g\beta \frac{{\partial \theta }}{{\partial X}} = \frac{1}{{4\pi \rho _0 }}J(\phi ,\nabla ^2 \phi ) - J(\psi ,\nabla ^2 \psi ) \end{equation}
\begin{equation} \label{eq23n} 
 \left(\frac{\partial}{\partial t}-\nu\nabla^2\right)v- 2\Omega_{00}f_R (1+\textrm{Ro})\frac{{\partial \psi }}{{\partial Z}}-\frac{{B_{00}f_m }}{{4\pi\rho_0}}\frac{{\partial \widetilde v}}{{\partial Z}}=\frac{1}{{4\pi\rho_0 }}J(\phi ,\widetilde v) - J(\psi ,v)
 \end{equation}
The notation $ J (a, b) = \frac {{\partial a}} {{\partial X}} \frac {{\partial b}} {{\partial Z}} - \frac {{\partial a}} {{\partial Z}} \frac {{\partial b}} {{\partial X}} $ - the Jacobian operator or the Poisson bracket $ J (a, b) \equiv \left\{a, b \right \} $.

Further, by differentiating the equation (\ref{eq17n}) with respect to $ Z $ and by differentiation  the equation (\ref{eq19n}) with respect to X,    and then by subtracting them from each other, we can find the equation for the current $ I $:
\begin{equation} \label{eq24n} 
\left(\frac{\partial}{\partial t}-\eta\nabla^2\right)I+\frac{\partial}{\partial Z}\left(u\frac{\partial \widetilde u}{\partial X}+w\frac{\partial \widetilde u}{\partial Z}- \widetilde u \frac{\partial u}{\partial X}- \widetilde w\frac{\partial u}{\partial Z}\right)-$$
$$-\frac{\partial}{\partial X}\left(u\frac{\partial \widetilde w}{\partial X}+w\frac{\partial \widetilde w}{\partial Z}- \widetilde u \frac{\partial w}{\partial X}- \widetilde w \frac{\partial w}{\partial Z}\right)=B_{00}f_m\frac{\partial\omega}{\partial Z}
\end{equation}
Equations (\ref{eq20n}) and (\ref {eq14n}) can also be written in a compact form using the definitions of the stream functions $ \psi $ and $ \phi $:
\begin{equation} \label{eq25n}
\left(\frac{\partial}{\partial t}-\eta\nabla^2\right)\phi - B_{00}f_m\frac{\partial\psi}{\partial Z}=-J(\psi,\phi)
\end{equation}
\begin{figure}
 \centering
\includegraphics[width=12 cm, height=8 cm]{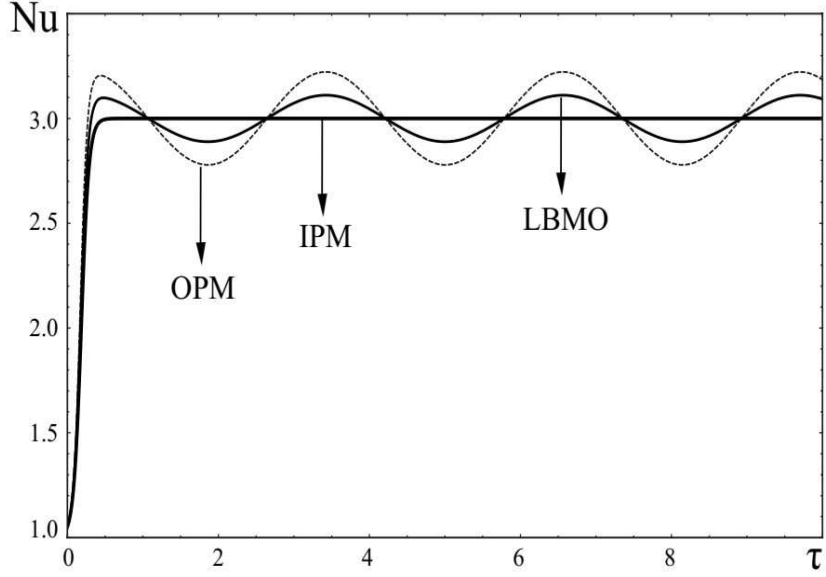}\\
\caption{ Three types of temperature modulation: $ \textrm {IPM}, \textrm {OPM}, \textrm {LBMO} $ for fixed parameters $ \textrm {Ro} = 2 $, $ \omega_T = 2 $, $ \delta_1 = 0.5 $.}\label{fg4t}
\end{figure}
\begin{equation} \label{eq26n}
 \left(\frac{\partial}{\partial t}-\eta\nabla^2\right)\widetilde v-B_{00}f_m \frac{{\partial v}}{{\partial Z}} + 2\Omega_{00}f_R\textrm{Ro}\frac{{\partial \phi }}{{\partial Z}}  = J(\phi ,v) - J(\psi,\widetilde v)\end{equation}
The form of the equation (\ref{eq16n}) for temperature disturbances is simplified in a similar way:
\begin{equation} \label{eq27n}
\left(\frac{\partial}{\partial t}-\chi\nabla^2\right)\theta-A\frac{\partial\psi}{\partial X}+\epsilon^2\delta_1\cdot\frac{\partial f_1}{\partial Z} \frac{\partial \psi}{\partial X}\triangle T=-J(\psi,\theta) \end{equation}
Equations (\ref{eq18n}), (\ref{eq19n}), (\ref{eq21n}), (\ref{eq22n}) and (\ref{eq23n}) together with the boundary conditions
\begin{equation} \label{eq28n}\psi=\nabla^2\psi=0,\quad \frac{dv}{dZ}=0,\quad \widetilde v=0, \quad \frac{d \phi}{dZ}=0,\quad \theta=0 \quad \textrm{at} \quad Z=0,h \end{equation}
describe nonuniformly rotating convection under the action of modulation of external fields.
For convenience, in equations (\ref{eq22n})-(\ref{eq27n})  we turn to the dimensionless variables, which we mark with an asterisk:
\[ (X,Z)=h(x^*,z^*), \; t=\frac{h^2}{\nu}t^*,\; \psi=\chi\psi^*,\; \phi=hB_{00} \phi^*,\; v=\frac{\chi}{h}v^*,\; \widetilde v=B_{00} \widetilde v^*,\; \theta=Ah\theta^* .\]
Omitting the asterisk symbol, we can rewrite equations (\ref{eq22n})-(\ref{eq27n}) in dimensionless variables:
\[\left(\frac{\partial }{{\partial t}}  - \nabla ^2\right)\nabla ^2 \psi  + \sqrt {\textrm{Ta}}\cdot f_R \frac{{\partial v}}{{\partial z}} - \Pr\textrm{Pm}^{-1}\textrm{ Q}f_m\frac{\partial }{{\partial z}}\nabla ^2 \phi  -\textrm{Ra}\cdot f_g\frac{{\partial \theta }}{{\partial x}}  =$$
$$= \Pr\textrm{Pm}^{-1}\textrm{Q}\cdot J(\phi ,\nabla ^2 \phi ) - {\Pr}^{-1}\cdot J(\psi ,\nabla ^2 \psi ) \]
\begin{figure}
 \centering
\includegraphics[width=12 cm, height=8 cm]{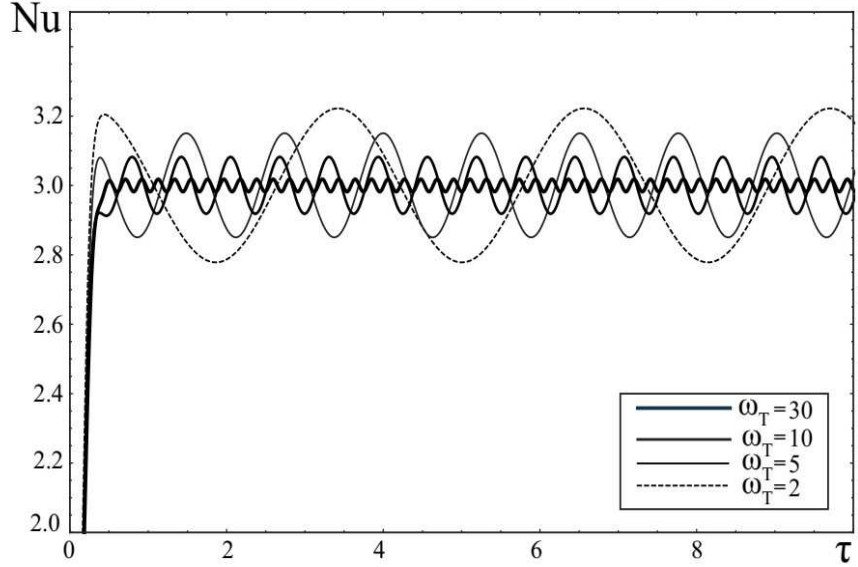} \\
\caption{Dependency of the Nusselt number $ \textrm {Nu} $ on the time $ \tau $ for the case of phase $ (\varphi = \pi) $ temperature modulation for  frequency $ \omega_T = (2, 5, 10, 30) $, amplitude $ \delta_1 = 0.5 $ and Rossby number $ \textrm {Ro} = 2 $.}\label{fg5t}
\end{figure}
\[\left(\frac{{\partial }}{{\partial t}}- \nabla ^2\right)v - \sqrt {\textrm{Ta}}\cdot f_R (1 +\textrm{Ro})\frac{{\partial \psi }}{{\partial z}} - \Pr\textrm{Pm}^{-1}\textrm{Q}f_m\frac{{\partial \widetilde v}}{{\partial z}}=$$
$$=\Pr\textrm{Pm}^{-1}\textrm{Q}\cdot J(\phi ,\widetilde v) - {\Pr}^{-1}\cdot J(\psi ,v) \]
\begin{equation} \label{eq29n} 
 \left(\frac{{\partial}}{{\partial t}}-\textrm{Pm}^{-1} \nabla^2\right)\phi-{\Pr}^{-1}f_m \frac{{\partial \psi }}{{\partial z}}=-{\Pr}^{-1}J(\psi ,\phi )) \end{equation}
\[\left(\frac{{\partial }}{{\partial t}}-\textrm{Pm}^{-1}\nabla^2\right)\widetilde v-{\Pr}^{-1}f_m \frac{{\partial v}}{{\partial z}} +\textrm{Ro}\sqrt{\textrm{Ta}}\cdot f_R \frac{{\partial \phi }}{{\partial z}} = {\Pr}^{-1}(J(\phi,v) - J(\psi,\widetilde v))\]
\[\left(\Pr\frac{{\partial}}{{\partial t}}- \nabla^2\right)\theta-\frac{{\partial \psi}}{{\partial x}}\left(1-\epsilon^2\delta_1 \frac{\partial f_1}{\partial z}\right)= - J(\psi,\theta ), \]
where the dimensionless parameters are: $ \textrm{Pr} = \nu / \chi $ ( Prandtl number), $ \textrm {Pm} = \nu / \eta $ (magnetic Prandtl number), $ \textrm {Ta} = \frac {4 \Omega_ {00}^2 h^4} {\nu^2} $ (Taylor number), $ \textrm {Q} = \frac {B_{00}^2 h^ 2} {4 \pi \rho_0 \nu \eta} $ (Chandrasekhar number), $ \textrm{Ra} = \frac {g_0 \beta A h^4} {\nu \chi} $ ( Rayleigh number on scale $ h $).

In the absence of the thermal phenomena $\textrm{Ra}=0$ and $f_R=f_m=f_g=1$, the system of equations (\ref{eq29n})  was used to study the nonlinear saturation mechanism of the standard MRI \cite{47s}.
In the case when the external field modulation is absent $ \delta_{1,2,3,4} = 0 $ and $ \textrm{Ra} \neq 0 $, the system of equations (\ref{eq29n}) was used to study the weakly nonlinear and chaotic modes of stationary convection in a nonuniformly rotating magnetoactive electrically conductive medium \cite {11s}-\cite{13s}.
\begin{figure}
 \centering
\includegraphics[width=12 cm, height=8 cm]{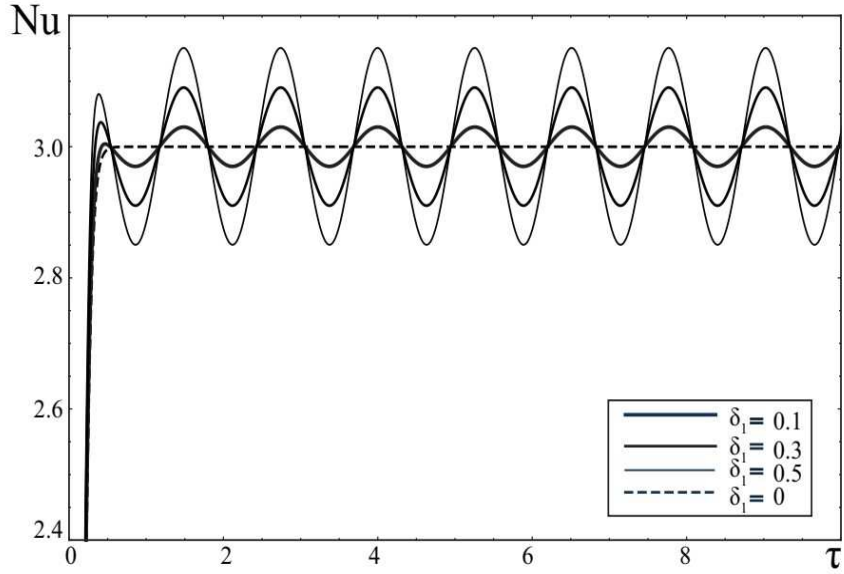}\\
\caption{Variations of the Nusselt number $ \textrm {Nu} $ depending on the amplitude of the phase temperature modulation $ \delta_1 = (0, 0.1, 0.3, 0.5) $ for the Rossby number $ \textrm {Ro} = 2 $ and the frequency $ \omega_T = 5 $.}\label{fg6t}
\end{figure}

\section{Equations of evolution of finite amplitude for different types of modulation}

In this section, we have analyzed the nonlinear stage of stationary convection in a nonuniformly rotating electrically conductive medium in a constant magnetic field under the influence of small oscillations: a) the temperature field at the layer boundaries; b) gravitational field; c) external magnetic field; g) the angular velocity of rotation. We  have considered all these effects separately (see Fig. \ref{fg2}).
Then we have compared the value of the heat transfer (Nusselt number) for each type of modulation, i.e. quantify heat transfer in terms of finite amplitudes. These amplitudes arise when an interaction occurs between several modes of perturbations. Such an interaction can be described only in the framework of a nonlinear or weakly nonlinear theory based on the perturbation theory method. Here we have performed our research in the framework of a weakly nonlinear theory. The small expansion parameter in this theory is the relative deviation of the Rayleigh number $ \textrm {Ra} $ from the critical value $ \textrm {Ra}_c $:
\[\epsilon^2=\frac{\textrm{Ra}-\textrm{Ra}_c}{\textrm{Ra}_c} \ll 1 \]
Then all the perturbed quantities $ {\bf{U}} $ in equations of the type $ \mathcal L {\bf {U}} = -N ({\bf {U}} | {\bf{U}})$ ($ N (\ldots) $ are nonlinear terms) are represented as a series in the perturbation theory
\[ {\bf{U}} \to \epsilon {\bf{U}}^{(1)}+\epsilon^2 {\bf{U}}^{(2)}+\epsilon^3 {\bf{U}}^{(3)}+\ldots \]
The equations for the perturbations in various orders of $\epsilon$ take the following form:
\[\epsilon^1: \mathcal L^{(0)} {\bf{U}}^{(1)}= 0,\]
\[\epsilon^2: \mathcal L^{(0)} {\bf{U}}^{(2)}= -N({\bf{U}}^{(1)}|{\bf{U}}^{(1)}) \]
\[\epsilon^3: \mathcal L^{(0)} {\bf{U}}^{(3)}= -\mathcal L^{(2)} {\bf{U}}^{(0)}-N({\bf{U}}^{(1)}|{\bf{U}}^{(2)})-N({\bf{U}}^{(2)}|{\bf{U}}^{(1)}) \]
The condition for solving this chain of nonlinear equations is known as Fredholm$^{,}$s alternative (see, for example \cite{48s} )
\begin{equation} \label{eq30n} 
 \left\langle {\bf{U}}^{\dagger}, R.H.\right\rangle =0 .
 \end{equation}
Here ${\bf{U}}^{\dagger}$ is a non-trivial solution of the linear self-adjoint problem $\mathcal L^{\dagger} {\bf{U}}^{\dagger}= 0 $, where   $\mathcal L^{\dagger} $  is a self-adjoint operator, which is determined from the following relation:
\begin{equation} \label{eq31n} 
 \left\langle {\bf{U}}^{\dagger}, \mathcal L {\bf{U}}\right\rangle \equiv \left\langle  \mathcal L^{\dagger}{\bf{U}}^{\dagger}, {\bf{U}}\right\rangle,  \end{equation}
where $\left\langle , \right\rangle$  is the inner product, which here has the following definition:
$$\left\langle {\bf{f}} , {\bf{g}}\right\rangle= \int\limits_{z=0}^1\int\limits_{x=0}^{2\pi/k_c} {\bf{f}}\cdot {\bf{g}}\, dxdz ,  $$ 
where $R.H.$ are right sides of the perturbed equations with nonlinear terms.
We represent all the variables in equations  (\ref{eq29n}) as an asymptotic expansion:
\[ \textrm{Ra}=\textrm{Ra}_c+\epsilon^2 \textrm{R}_{2}+\epsilon^4 \textrm{R}_{4}+\ldots \]
\[\psi=\epsilon \psi_1+\epsilon^2 \psi_2+\epsilon^3\psi_3+\ldots\]
\begin{equation} \label{eq32n}
 v=\epsilon v_1+\epsilon^2 v_2+\epsilon^3 v_3+\ldots  \end{equation}
\[\phi=\epsilon \phi_1+\epsilon^2 \phi_2+\epsilon^3\phi_3+\ldots\]
\[\widetilde v=\epsilon \widetilde v_1+\epsilon^2 \widetilde v_2+\epsilon^3\widetilde v_3+\ldots\]
\[\theta=\epsilon \theta_1+\epsilon^2 \theta_2+\epsilon^3\theta_3+\ldots\]
Here $ \textrm{Ra}_c$ is the critical value of the Rayleigh number for convection without modulation. The amplitudes of the perturbed quantities depend only on the slow time $\tau=\epsilon^2 t $. For simplicity let us take into account the nonlinear terms in (\ref{eq29n}) only in the heat
balance equation. As it is shown in \cite{23s}, this approximation is equivalent to applying the Galerkin approximation of the minimum order to the equations (\ref{eq29n}).
In the lowest order, we get the equation:
\begin{equation} \label{eq33n} \widehat{L}M_1=0, \end{equation}
\begin{figure}
  \centering
	\includegraphics[width=13 cm, height=8 cm]{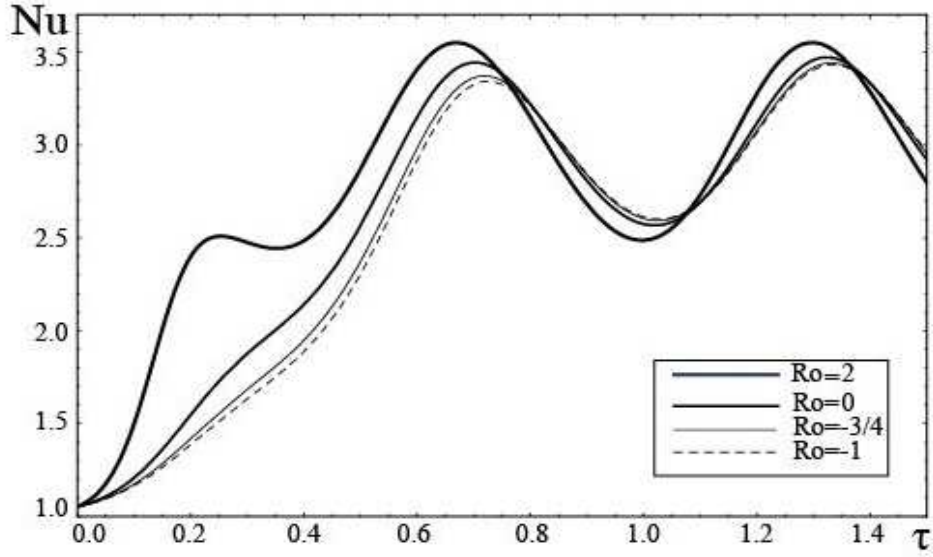}   \\
	\caption{Dependency of the Nusselt number $ \textrm{Nu} $ on the time $ \tau $ for Rossby numbers $ \textrm {Ro} = (2, 0, -3/4, -1) $ in an oscillating gravitational field with a frequency $ \omega_g = $ 10 and an amplitude $ \delta_4 = 0.3 $. }\label{fg3g}
	\end{figure}
where $M_1=\begin{bmatrix} \psi _1 \\ v_1 \\ \phi _1 \\ \widetilde v_1  \\ \theta _1 \\ \end{bmatrix}$, $\widehat{L}$ is the matrix operator of the form:
\[\widehat L =\begin{bmatrix}
    - \nabla ^4  & \sqrt {\textrm{Ta}}\frac{\partial }{\partial z} &  - \textrm{Pm}^{ - 1} \Pr\textrm{Q}\frac{\partial }{\partial z}\nabla ^2  & 0 & - \textrm{Ra}_c \frac{\partial }{\partial x}  \\ 
		
		- \sqrt {\textrm{Ta}}(1 +\textrm{Ro})\frac{\partial }{\partial z} & - \nabla ^2  & 0 &  -\textrm{Pm}^{ - 1} \Pr\textrm{Q}\frac{\partial }{\partial z} & 0  \\
- \Pr ^{-1} \frac{\partial }{\partial z} & 0 &  - \textrm{Pm}^{-1} \nabla ^2  & 0 & 0  \\

0 & - \Pr ^{-1} \frac{\partial }{\partial z} &  \textrm{Ro}\sqrt{\textrm{Ta}}\frac{\partial }{\partial z}  & - \textrm{Pm}^{-1} \nabla ^2 & 0  \\
		
 -{\Pr}^{-1} \frac{\partial}{\partial x} & 0 & 0 &  0 &  -{\Pr}^{-1} \nabla ^2   \\

\end{bmatrix} . \]
The solutions of the system of equations (\ref{eq33n}) with the boundary conditions of (\ref{eq28n}) have, respectively, the following  form:
\[\psi_1=A(\tau)\sin k_c x \sin \pi z,\; \theta_1=\frac{A(\tau)k_c}{a^2}\cos k_c x \sin \pi z,\; \phi_1=\frac{A(\tau)\pi \textrm{Pm}}{a^2 \Pr}\sin k_c x \cos \pi z ,   \]
\begin{equation} \label{eq34n}
\widetilde v_1=-\frac{A(\tau)\pi^2\sqrt{\textrm{Ta}}(1+\textrm{Ro}(1-\textrm{Pm}))\textrm{Pm}}{\Pr(a^4+\pi^2\textrm{Q})}\sin k_c x \sin \pi z,
\end{equation}
\[ v_1=\frac{A(\tau)\pi \sqrt{\textrm{Ta}}}{a^2}\cdot\frac{(1+\textrm{Ro})a^4+\pi^2 \textrm{QPmRo}}{a^4+\pi^2 \textrm{Q}}\sin k_c x \cos \pi z,\quad a^2=k_c^2+\pi^2. \]
The amplitude $A(\tau)$ is still unknown. The critical value of the Rayleigh number  $\textrm{Ra}_c$  for the stationary magnetoconvection in a nonuniformly rotating electrically conducting medium is found from the first equation of the system (\ref{eq34n}) and has the form of the formula obtained in the linear theory \cite{11s}-\cite{12s}:
\begin{figure}
  \centering
\includegraphics[width=16 cm, height=21 cm]{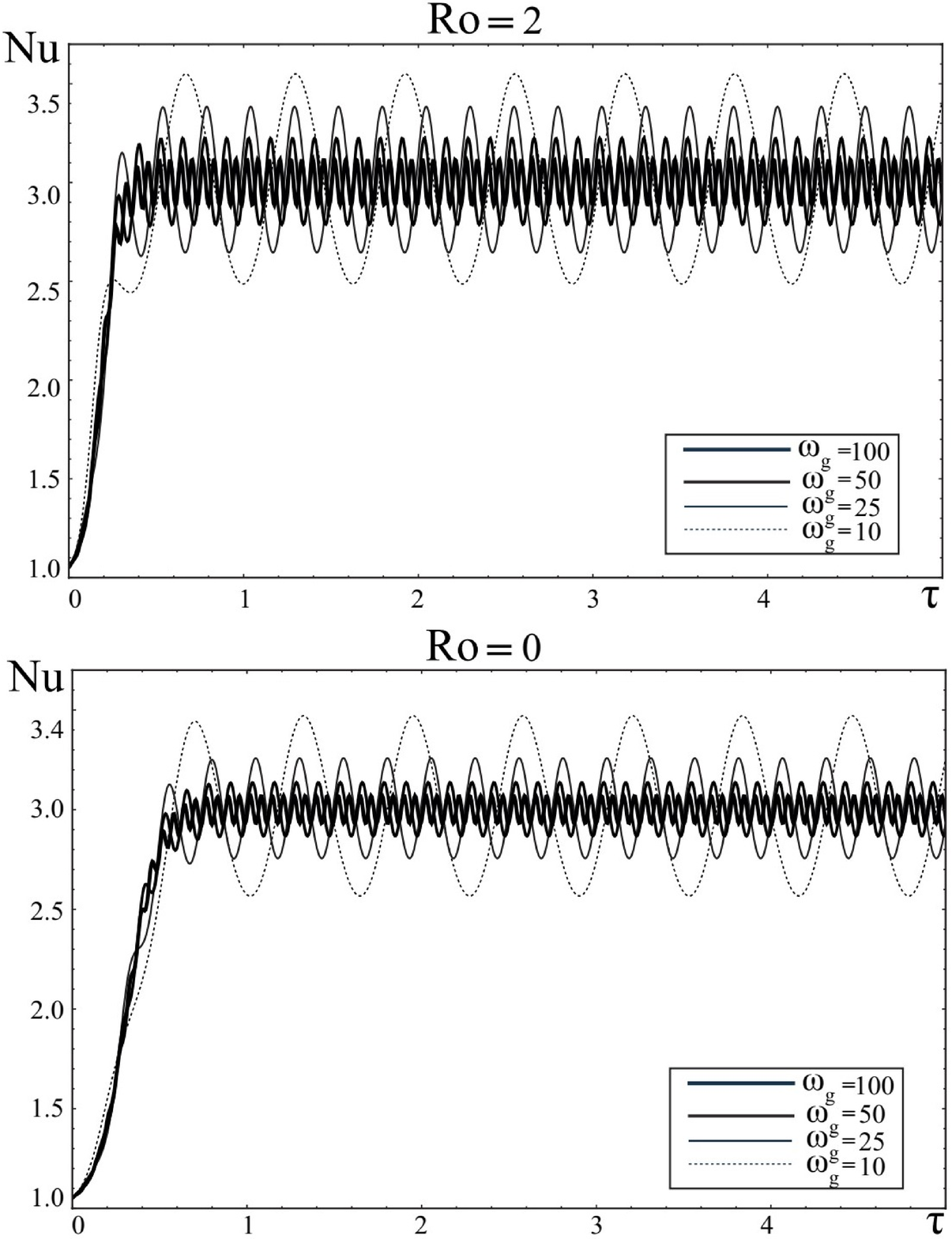}   \\
	\caption{ Dependency of the Nusselt number $ \textrm{Nu} $ on the time $ \tau $ for positive Rossby numbers $ \textrm{Ro} = (2, 0) $ in an oscillating gravitational field with a frequency $ \omega_g = (10, 25, 50, 100) $ and an amplitude $ \delta_4 = 0.3 $. }\label{fg4g}
	\end{figure}
	\begin{figure}
  \centering
	\includegraphics[width=16 cm, height=21 cm]{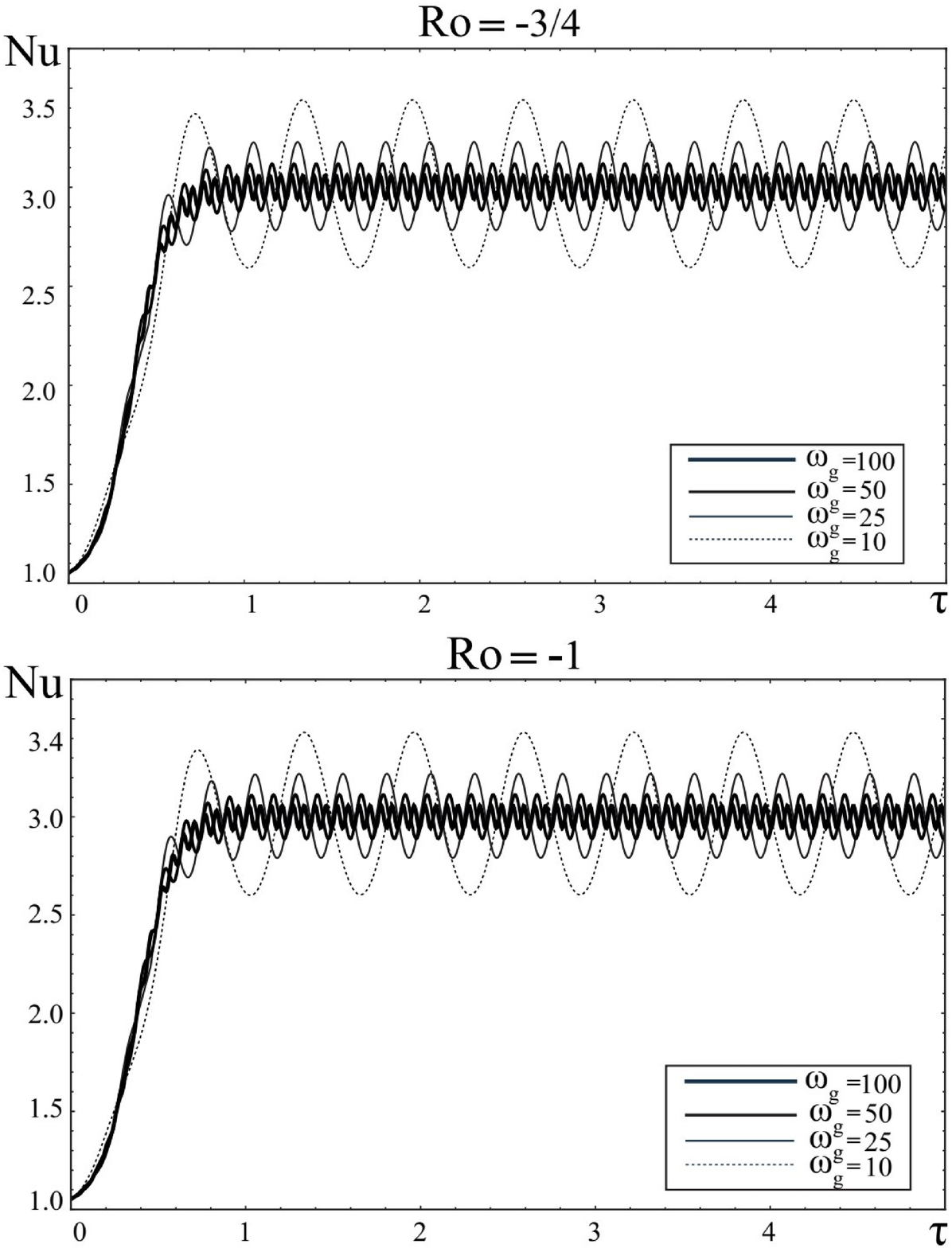}   \\
	\caption{Dependency of the Nusselt number $ \textrm{Nu} $ on time $ \tau $ for negative Rossby numbers $ \textrm{Ro} = (- 3/4, -1) $ in an oscillating gravitational field with a frequency $ \omega_g = (10, 25, 50, 100) $ and an amplitude $ \delta_4 = 0.3 $. }\label{fg4gr}
	\end{figure}
	\begin{equation} \label{eq35n}
  \textrm{Ra}_{c}  = \frac{{(\pi ^2  + k_c^2 )^3 }}{{k_c^2 }} + \frac{{\pi ^2 (\pi ^2  + k_c^2 ) \textrm{Q}}}{{k_c^2 }} + \frac{{\pi ^2 (\pi ^2  + k_c^2 )^2 \textrm{Ta}}}{{k_c^2 ((\pi ^2  + k_c^2 )^2  + \pi ^2 \textrm{Q})}} + \frac{{\pi ^2  \textrm{TaRo}((\pi ^2  + k_c^2 )^2  + \pi ^2 \textrm{QPm})}}{{k_c^2((\pi ^2  + k_c^2 )^2  + \pi ^2 \textrm{Q}) }}.\end{equation}
It should be noted that for the absence of  heating $ \textrm {Ra} = 0 $, the threshold value of the hydrodynamic Rossby number $ \textrm {Ro} $ has the form:
\[\textrm{Ro}_{\textrm{cr}}  =-\frac{a^2(a^4+\pi^2\textrm{Q} )^2+\pi^2a^4 \textrm{Ta}}{\pi^2\textrm{Ta}(a^4+\pi^2 \textrm{Q} \textrm{Pm})}.\]
Passing to dimensional variables
\[\frac{\pi^2 \textrm{Q}}{a^4} \rightarrow \frac{\omega_A ^2}{\omega_\nu \omega_\eta},\; \frac{\pi^2\textrm{Q} \textrm{Pm}}{a^4} \rightarrow \frac{\omega_A ^2}{ \omega_\eta ^2},\; \frac{\textrm{Ta}}{a^4} \rightarrow \frac{4\Omega^2}{\omega_\nu^2},\; \frac{\pi^2}{a^2}\rightarrow \xi^2\]
we obtain the expression for $\textrm{Ro}_{\textrm{cr}}$ \cite{41s}:
\[\textrm{Ro}_{\textrm{cr}}=-\frac{{(\omega _A^2+\omega_\nu\omega_\eta)^2+4\xi^2\Omega^2 \omega_\eta^2 }}{{4\Omega^2 \xi^2 (\omega_A^2 +\omega_\eta^2 )}},  \]
where the following notation has been introduced: $\omega_{\nu} =\nu k^{2} $ and $\omega_{\eta }=\eta k^{2}$ are the viscous and Ohmic frequencies, respectively, and $\omega_{A} $ is the Alfven frequency, $\omega_{A}^{2} =k_{z}^{2} c_{A}^{2} =\frac{k_{z}^{2} B_{0}^{2} }{4\pi \rho _{0} }$.
Therefore, in the limiting case of $\textrm{Ra}=0$, magnetorotational instability appears in a nonuniformly rotating electroconducting fluid in a constant magnetic field. 
\begin{figure}
  \centering
	\includegraphics[width=13 cm, height=9 cm]{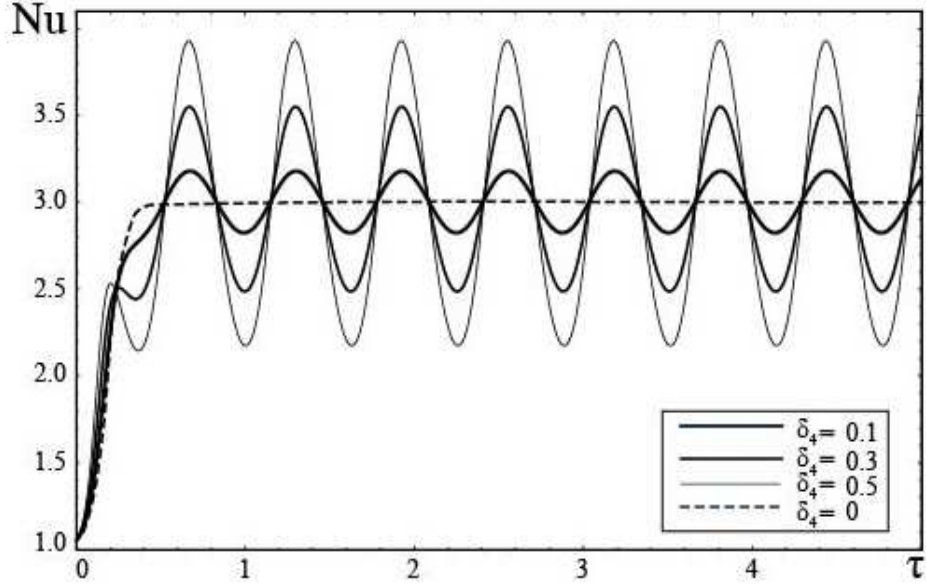}   \\
	\caption{ Variations of the Nusselt number $ \textrm {Nu} $ depending on the amplitude of the oscillating gravitational field $ \delta_4 = (0, 0.1, 0.3, 0.5) $ for the Rossby number $ \textrm {Ro} = 2 $ and the frequency $ \omega_g = 10 $.}\label{fg5g}
	\end{figure}
The criterion for its appearance is the condition imposed on the angular velocity profile $\Omega(R)$ of the rotating liquid, i.e., Rossby number $\textrm{Ro}>\textrm{Ro}_{\textrm{cr}}$. 
Figure \ref{fg3} shows diagrams of the dependency of the critical Rayleigh number  $\textrm{Ra}_{c}$ on the wavenumbers for various angular velocity profiles (Rossby numbers $\textrm{Ro} $).  It can be seen that for negative Rossby numbers $ \textrm {Ro} <0 $ the critical Rayleigh number $ \textrm{Ra}^{\textrm{min}}_c $ becomes smaller than in the case of uniform rotation $ \textrm{Ro} = 0 $ and rotation with positive numbers $ \textrm {Ro}> 0 $.

According to the formula (\ref{eq31n}), it is necessary to find solutions of the linear self-adjoint problem $\widehat L^{\dagger} M_1^{\dagger}=0$, where the matrix $ M_1^{\dagger}$ has the form: $M_1^{\dagger} =(\psi_1^{\dagger},\theta_1^{\dagger},\phi_1^{\dagger},v_1^{\dagger})^{Tr}$ and $\widehat L^{\dagger}$ is a self-adjoint matrix operator:
\begin{equation} \label{eq36n} 
\widehat L^{\dagger} =\begin{bmatrix}
    - \nabla ^4  & \textrm{Ra}_c\widehat P\frac{\partial }{\partial x} &   \textrm{Pm}^{-1}\Pr\textrm{Q}\widehat P\frac{\partial }{\partial z}\nabla^2  & -\sqrt {\textrm{Ta}}\widehat P\frac{\partial }{\partial z}  \\ 
		 
		\textrm{Ra}_c\widehat P\frac{\partial }{\partial x}  &  -\textrm{Ra}_c\widehat P\nabla^2  & 0 & 0  \\
		
		\textrm{Pm}^{-1}\Pr\textrm{Q}\widehat P\frac{\partial }{\partial z}\nabla^2 & 0 &  -\textrm{Q}\Pr^{2}\textrm{Pm}^{-2}\widehat P\nabla^4  & 0   \\

    -\sqrt {\textrm{Ta}}\widehat P\frac{\partial }{\partial z}  & 0 &  0  & \left(\nabla^4-\textrm{Q} \frac{\partial^2 }{\partial z^2}\right)\nabla^2 \\
	\end{bmatrix} 
  \end{equation}
The solutions of the system of equations (\ref{eq36n}) have the form:
\[\psi_1^{\dagger}=A(\tau)\sin k_c x \sin \pi z,\]
\begin{equation} \label{eq37n}
 \theta_1^{\dagger}=-\frac{A(\tau)k_c}{a^2}\cos k_c x \sin \pi z,\end{equation}
\[ \phi_1^{\dagger}=-\frac{A(\tau)\pi \textrm{Pm}}{a^2 \Pr}\sin k_c x \cos \pi z ,\]
\begin{figure}
  \centering
	\includegraphics[width=15 cm, height=20 cm]{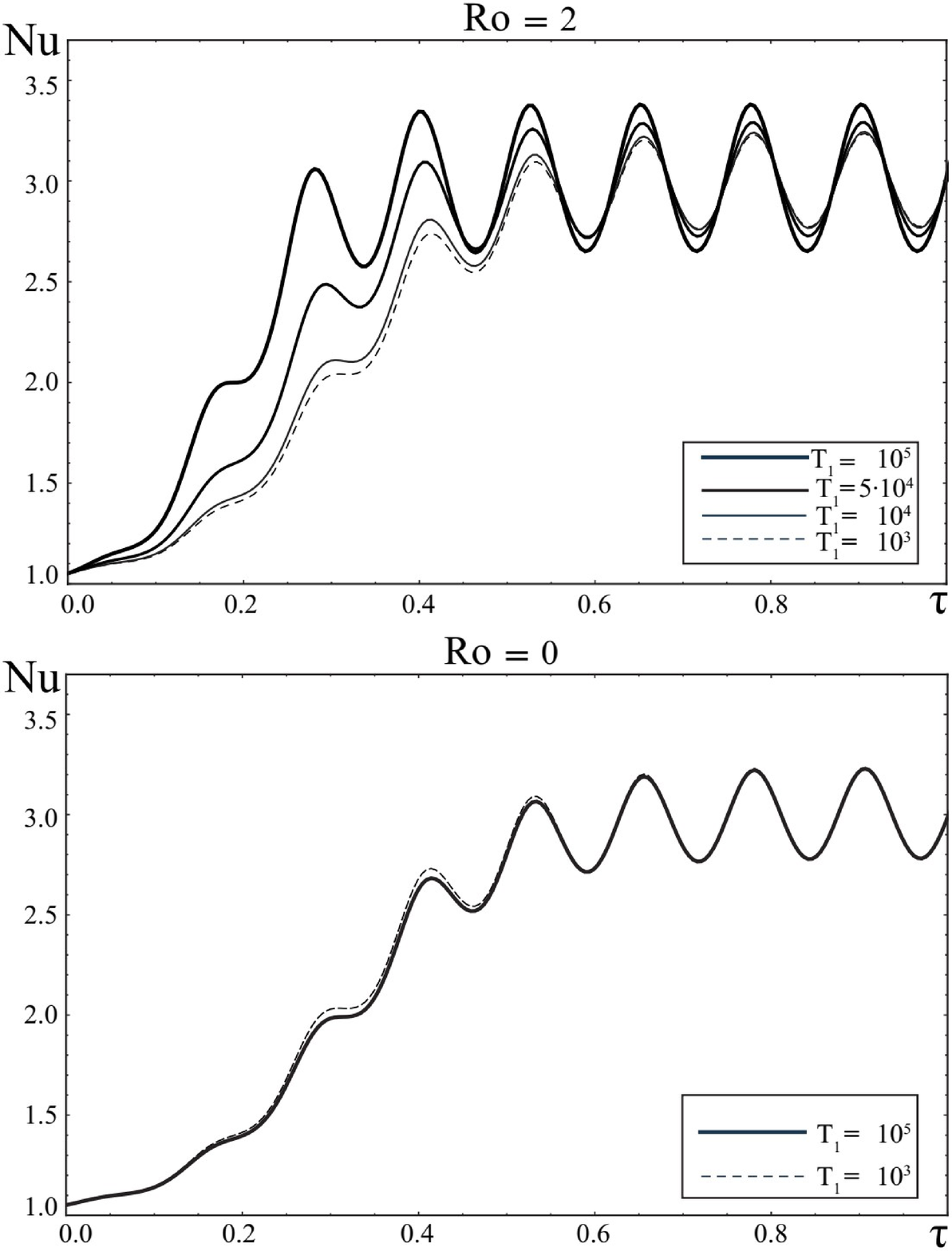}   \\
	\caption{Variations of the Nusselt number $ \textrm{Nu} $ depending on the Taylor number $ \textrm{T}_1 = (10^3, 10^4, 5 \cdot 10^4, 10^5) $ for the amplitude parameters of the oscillating gravitational field $ \delta_4 = 0.5 $, frequencies $ \omega_g = 50 $ for positive Rossby numbers $ \textrm{Ro} = (2, 0) $. }\label{fg6g}
	\end{figure}
	\begin{figure}
  \centering
	 \includegraphics[width=15 cm, height=20 cm]{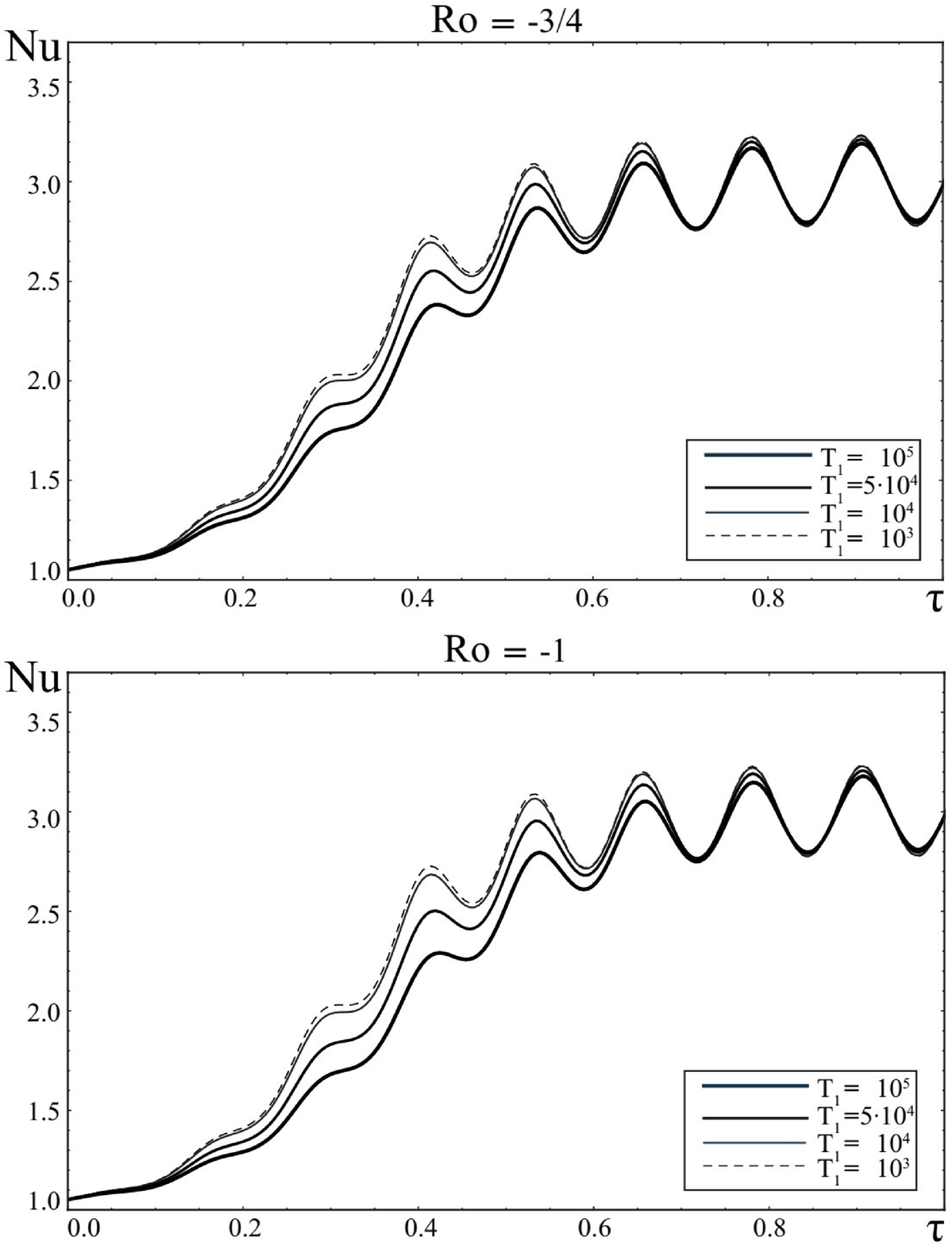}  \\
	\caption{ Variations of the Nusselt number $ \textrm {Nu} $ depending on the Taylor number $ \textrm {T}_1 = (10^3, 10^4, 5 \cdot 10^4, 10^5) $ for the amplitude parameters of the oscillating gravitational field $ \delta_4 = 0.5 $, frequencies $ \omega_g = 50 $ for negative Rossby numbers $ \textrm {Ro} = (- 3/4, -1) $.}\label{fg6gr}
	\end{figure}
\[v_1^{\dagger}=-\frac{A(\tau)\pi \sqrt{\textrm{Ta}}}{a^2}\cdot\frac{(1+\textrm{Ro})a^4+\pi^2 \textrm{QPmRo}}{a^4+\pi^2 \textrm{Q}}\sin k_c x \cos \pi z .\]
For the second order of $\epsilon$, we have the following equation:
\begin{equation} \label{eq38n} \widehat{L}M_2=N_2, \end{equation}
where $M_2=\begin{bmatrix} \psi _2 \\ v_2 \\ \phi _2 \\ \widetilde v_2  \\ \theta _2 \\ \end{bmatrix}$, $N_2=\begin{bmatrix} N_{21} \\ N_{22} \\ N_{23} \\ N_{24}  \\ N_{25} \\ \end{bmatrix}$; 
\[N_{21}=N_{22}=N_{23}=N_{24}=0,\]
\[N_{25}=-{\Pr}^{-1}\left[\frac{\partial\psi_1}{\partial x}\frac{\partial \theta_1}{\partial z}-\frac{\partial \theta_1}{\partial x}\frac{\partial\psi_1}{\partial z}\right].\]
Using solutions of (\ref{eq34n}) and boundary conditions of (\ref{eq28n}), we can find solutions of equations (\ref{eq38n}):
\begin{equation} \label{eq39n}
\psi_2=0,\quad \theta_2=-\frac{A^2(\tau)k_c^2}{8\pi a^2}\sin(2\pi z),\quad \phi_2=0,$$ 
$$ v_2=0,\quad \widetilde v_2=0. \end{equation}
To analyze the intensity of the heat transfer, a horizontally-averaged heat flux is introduced at the boundary of the layer of electrically conducting fluid (Nusselt number):
\begin{equation} \label{eq40n}
\textrm{Nu}(\tau)=1+\frac{\left[\frac{k_c}{2\pi}\int\limits_0^{2\pi/k_c}\left(\frac{\partial \theta_2}{\partial z}\right)dx\right]_{z=0}}{\left[\frac{k_c}{2\pi}\int\limits_0^{2\pi/k_c}\left(\frac{\partial T_0}{\partial z}\right)dx\right]_{z=0}}=1+\frac{k_c^2}{4a^2}A^2(\tau)
 \end{equation}
The heat flow intensity (of Nusselt number  $\textrm{Nu}$) have been analyzed after the expression for the amplitude $A(\tau)$ is obtained.
As can be seen from the an asymptotic expansion (\ref {eq32n}), modulation effects contribute only in the third order in $ \epsilon $, so we have considered these effects separately from each other in the third order in $ \epsilon $.

\subsection{Temperature modulation of fluid layer boundaries}

Let us consider only the temperature modulation of the layer boundaries, then in the equations (\ref{eq29n}) it is necessary to put $ f_R = f_m = f_g = 1 $. At the third order, we have
\begin{equation} \label{eq41n} \widehat{L}M_3=N_3, \end{equation}
where $M_3=\begin{bmatrix} \psi _3 \\ v_3 \\ \phi _3 \\ \widetilde v_3  \\ \theta _3 \\ \end{bmatrix}$, $N_3=\begin{bmatrix} N_{31} \\ N_{32} \\ N_{33} \\ N_{34}  \\ N_{35} \\ \end{bmatrix}$; 
\[N_{31}=-\frac{\partial}{\partial\tau}\nabla^2\psi_1+\textrm{Ra}_2 \frac{\partial\theta_1}{\partial x}=\left(a^2\frac{\partial A(\tau)}{\partial \tau}-\textrm{Ra}_2 \frac{k_c^2 A(\tau)}{a^2}\right)\sin k_c x \sin \pi z,   \]
\begin{figure}
  \centering
	\includegraphics[width=12 cm, height= 8 cm]{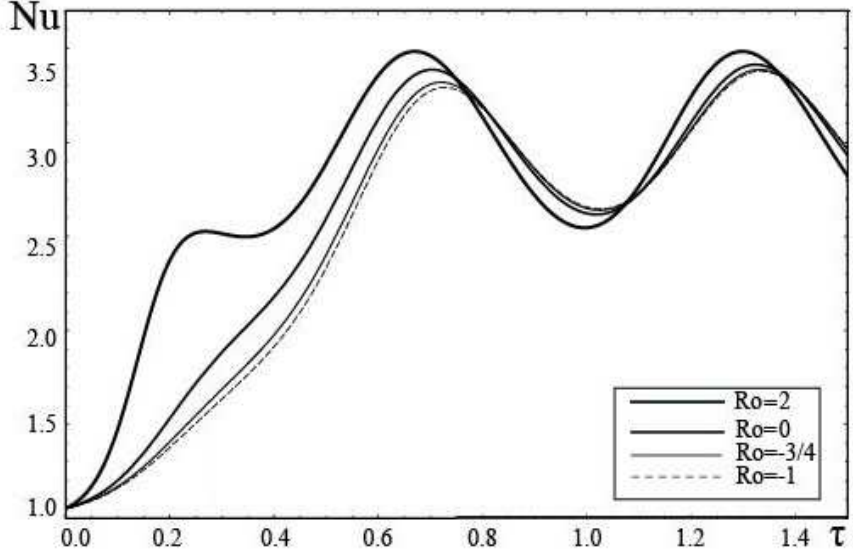}   \\
	\caption{Dependency of the Nusselt number $ \textrm{Nu} $ on the time $ \tau $ for Rossby numbers $ \textrm{Ro} = (2, 0, -3/4, -1) $ in an oscillating magnetic field with a frequency of $ \omega_B = $ 10 and with an amplitude of $ \delta_3 = 0.3 $. }\label{fg3m}
	\end{figure}
\[N_{32}=-\frac{\partial v_1}{\partial\tau}=-\frac{\pi \sqrt{\textrm{Ta}}}{a^2}\cdot\frac{(1+\textrm{Ro})a^4+\pi^2 \textrm{QPmRo}}{a^4+\pi^2 \textrm{Q}}\cdot\frac{\partial A(\tau)}{\partial \tau}\sin k_c x \cos \pi z, \]
\[N_{33}=-\frac{\partial\phi_1}{\partial\tau}=-\frac{\pi \textrm{Pm}}{a^2 \Pr}\cdot\frac{\partial A(\tau)}{\partial \tau}\sin k_c x \cos \pi z    \]
\[N_{34}=-\frac{\partial \widetilde v_1}{\partial\tau}= \frac{\pi^2\sqrt{\textrm{Ta}}(1+\textrm{Ro}(1-\textrm{Pm}))\textrm{Pm}}{\Pr(a^4+\pi^2\textrm{Q})}\cdot\frac{\partial A(\tau)}{\partial \tau}\sin k_c x \sin \pi z , \]
\[N_{35}=-\frac{\partial\theta_1}{\partial\tau}-{\Pr}^{-1}\delta_1\frac{\partial f_1}{\partial z}\frac{\partial \psi_1}{\partial x}-{\Pr}^{-1}\left[\frac{\partial\psi_1}{\partial x}\frac{\partial\theta_2}{\partial z}-\frac{\partial\theta_2}{\partial x}\frac{\partial\psi_1}{\partial z} + \frac{\partial\psi_2}{\partial x}\frac{\partial\theta_1}{\partial z}-\frac{\partial\theta_1}{\partial x}\frac{\partial\psi_2}{\partial z} \right]=\]
\[ =-\frac{k_c}{a^2}\frac{\partial A(\tau)}{\partial \tau}\cos k_c x \sin \pi z-{\Pr}^{-1}\delta_1\frac{\partial f_1}{\partial z}k_c A(\tau)\cos k_c x \sin \pi z+\frac{{\Pr}^{-1}k_c^3}{4a^2} A^3(\tau)\cos k_c x \sin \pi z\cos 2\pi z. \]
The solvability condition (Fredholm alternative) for the third-order equations $ O (\epsilon^3) $ is found from the formula (\ref{eq30n}):
\begin{equation} \label{eq42n} 
\int\limits_{z=0}^1\int\limits_{x=0}^{2\pi/k_c}\left[\widehat P \psi_1^{\dagger}\cdot {\mathcal R}_{31}+\textrm{Ra}_c \Pr\widehat P \theta_1^{\dagger}\cdot {\mathcal R}_{32}+\textrm{Q}\textrm{Pr}^2\textrm{Pm}^{-1}\widehat P \nabla^2\phi_1^{\dagger}\cdot {\mathcal R}_{33}+v_1^{\dagger}\cdot {\mathcal R}_{34}\right]dxdz=0,
\end{equation}
 where the notations are introduced  $$\widehat P=(1+\textrm{Ro})\nabla^4-\textrm{QPmRo}\frac{\partial^2}{\partial z^2}, \quad {\mathcal R}_{31}=N_{31},\quad {\mathcal R}_{32}=N_{35},\quad {\mathcal R}_{33}=N_{33},  $$
\[{\mathcal R}_{34}=-\nabla^4 N_{32}+\textrm{QPr}\frac{\partial}{\partial z}\nabla^2 N_{34}+\sqrt{\textrm{Ta}}\textrm{QPmPrRo}\frac{\partial^2 N_{33}}{\partial z^2} . \]
\begin{figure}
  \centering
	\includegraphics[width=15 cm, height=20 cm]{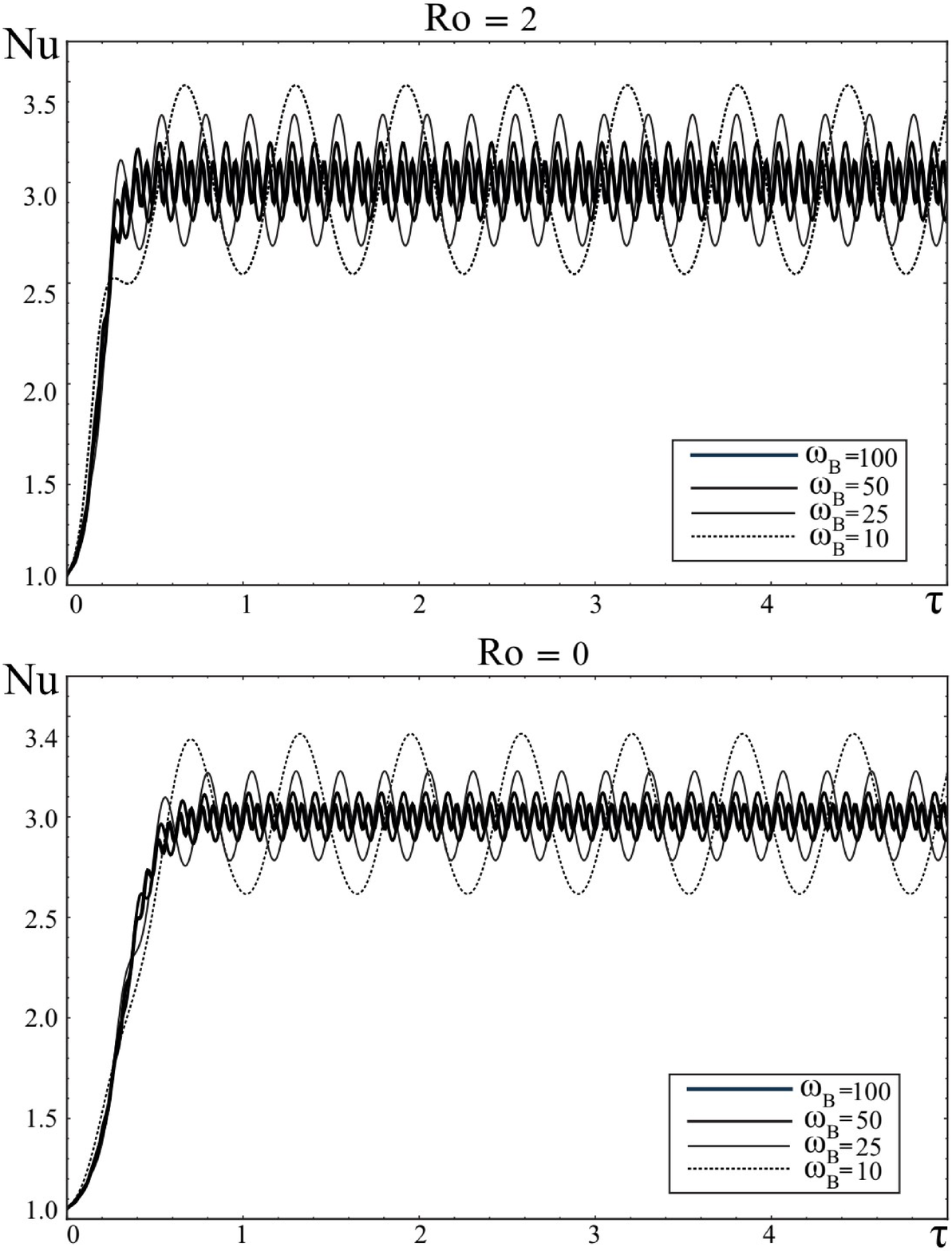}   \\
	\caption{Dependency of the Nusselt number $ \textrm{Nu} $ on the time $ \tau $ for positive Rossby numbers $ \textrm{Ro} = (2, 0) $ in an oscillating magnetic field with a frequency of $ \omega_B = (10, 25, 50, 100) $ and with an amplitude of $ \delta_3 = 0.3 $. }\label{fg4m}
	\end{figure}
	\begin{figure}
  \centering
	\includegraphics[width=15 cm, height=20 cm]{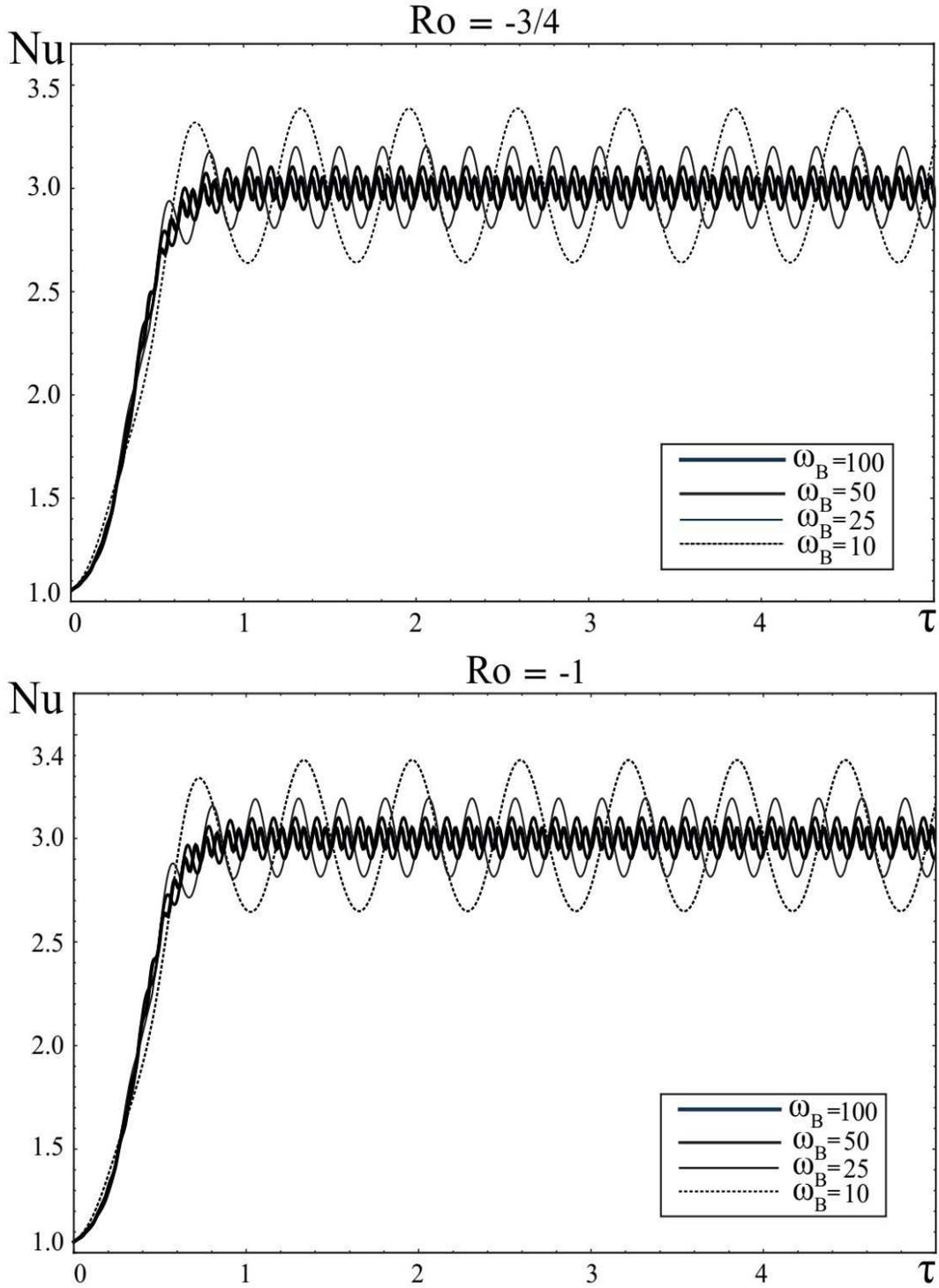}   \\
\caption{Dependency of the Nusselt number $ \textrm{Nu} $ on the time $\tau$ for negative Rossby numbers $ \textrm {Ro} = (- 3/4, -1) $ in an oscillating magnetic field with a frequency $ \omega_B = (10, 25, 50, 100) $ and with an amplitude of $ \delta_3 = 0.3 $.}\label{fg4mr}
\end{figure}
By integrating into (\ref{eq42n}), we obtained a nonlinear equation for the amplitude $ A (\tau) $, which refers to the non-autonomous Ginzburg-Landau (GL) equation for stationary convective instability, with a time-periodic coefficient in the following form:
\begin{equation} \label{eq43n} 
\mathscr A_{1T}\frac{\partial A}{\partial\tau}-\mathscr A_{2T}(\tau) A+\mathscr A_{3T} A^3=0
 \end{equation}
Here the coefficients $ \mathscr A_{1T, 2T, 3T} $ have the following form:
\begin{equation} \label{eq44n} 
\mathscr A_{1T}=a^2+\frac{k_c^2}{a^4}\Pr\textrm{Ra}_c-\frac{\pi^2}{a^2}\textrm{QPm}-\frac{\pi^2\textrm{Ta}\left((1+\textrm{Ro})a^4+\pi^2\textrm{QPm}(\textrm{RoPm}-1)\right)}{(a^4+\pi^2\textrm{Q})^2}-\frac{\pi^4\textrm{TaRoQ}\textrm{Pm}^2}{a^4(a^4+\pi^2\textrm{Q})}, $$
$$ \mathscr A_{2T}(\tau)=\frac{k_c^2 }{a^2}\textrm{Ra}_2-\frac{k_c^2\textrm{Ra}_c}{2a^2}\delta_1\cdot I(\tau),\quad \mathscr A_{3T}=\frac{k_c^4\textrm{Ra}_c}{8a^4}, \end{equation}
where the integral $ I (\tau) $ is equal to:
$$I(\tau)=\int\limits_0^1 \frac{\partial f_1(\tau,z)}{\partial z}\sin^2 \pi z dz=\textrm{Re}\left\{\frac{\pi^2 e^{-i\omega_T \tau}(e^{-i\varphi}-1)}{\lambda^2+4\pi^2}\right\}, \quad \omega_T= \frac{\widetilde{\omega}_T}{\epsilon^2}.$$
In the limiting case, when there is no temperature modulation $ \delta_1 = 0 $, the equation (\ref{eq44n}) was obtained in \cite{13s}. In the absence of modulation, the equation (\ref{eq44n}) has an analytical solution with the known initial condition $ A_0 = A (0) $:
\begin{equation} \label{eq45n}   A(\tau)=\frac{A_0}{\sqrt{\frac{\mathscr A_3}{\mathscr A_2}A_0^2+\left(1-A_0^2\frac{\mathscr A_3}{\mathscr A_2}\right)\exp\left(-\frac{2\tau\mathscr A_2}{\mathscr A_1}\right)}}
 \end{equation}
\begin{figure}
  \centering
	\includegraphics[width=12 cm, height=8 cm]{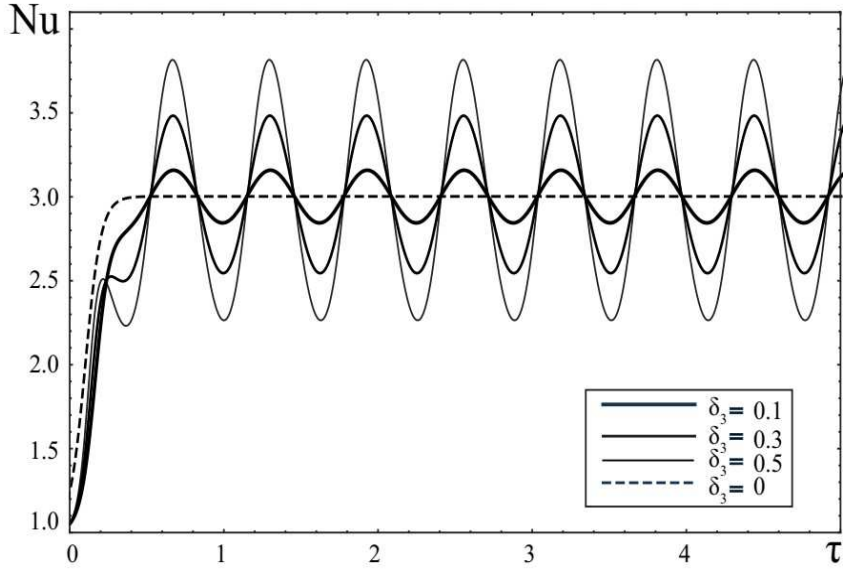}   \\
	\caption{Variations of the Nusselt number $ \textrm{Nu} $ depending on the amplitude of the oscillating magnetic field $ \delta_3 = (0, 0.1, 0.3, 0.5) $ for the Rossby number $ \textrm{Ro} = 2 $ and frequency $ \omega_B = 10 $. }\label{fg5m}
	\end{figure}
In \cite{23s}, the effect of different rotation profiles (Rossby numbers $\textrm{Ro}$) on the heat transfer value (Nusselt number $ \textrm{Nu}$) using (\ref{eq45n}) was studied.
The equation (\ref{eq44n}) have been solved numerically for different rotation profiles and modulation types:
\begin{enumerate}
\item In-phase modulation $(\textrm{IPM})$ $\varphi=0$,
\item Out-phase modulation $(\textrm{OPM})$ $\varphi=\pi$,
\item Only Lower boundary modulated $(\textrm{LBMO})$ $\varphi=-i\infty$,
which means that the modulation effect is not  considered in the upper boundary but only in the lower boundary.
\end{enumerate}
Further, we have considered $ \textrm{Ra}_2 \approx \textrm{Ra}_c $, since the nonlinearity is considered near the critical state of convection.

The results of numerical solutions of the equation (\ref{eq44n}) for the case of in-phase modulation $ \varphi = 0 $ for Rossby numbers $ \textrm {Ro} = (2, 0, -3/4, -1) $ are shown in Fig. \ref{fg3t}. 
The fixed parameters of convection and temperature modulation are respectively equal to:
$$\textrm{Q}/\pi^2=\textrm{Q}_1=80, \; \textrm{Ta}/\pi^4=\textrm{T}_1=10^5,\;\textrm{Ra}_c/\pi^4=\textrm{R}_1=9500,\; \textrm{Pm}=1,$$ 
$$\textrm{Pr}=10,\; A_0=0.5, \; \omega_T=2, \; \delta_1=0.5 .  $$
As can be seen from Fig. \ref{fg3t} the heat flow increases towards positive Rossby numbers $ (\textrm {Ro}> 0) $ in the time interval  $ \tau \in [0,1] $. The heat flow reaches the final value for all rotation profiles. 
Fig. \ref{fg3tf} and Fig. \ref{fg3tfm} show the numerical solutions of the equation (\ref{eq44n}) respectively for the case of $(\textrm{OPM})$ and 
$(\textrm{LBMO})$. Here we also see an increase in heat flux towards positive Rossby numbers $ (\textrm {Ro}> 0) $, then there is a periodic change in heat transfer for all rotation profiles. The comparison of results of in-phase modulation  $(\textrm{IPM})$, out of phase modulation $(\textrm{OPM})$ and when only lower boundary temperature is modulated $(\textrm{LBMO})$ for fixed parameters $ \textrm {Ro} = 2 $, $ \omega_T = 2 $, $ \delta_1 = 0.5 $ is presented  in Fig. \ref{fg4t}. Comparing these graphs, we can conclude that the variations in the Nusselt number $ \Delta \textrm {Nu} $ are greater for the case of out-phase modulation ($(\textrm{OPM})$): $$ \Delta\textrm{Nu}|_{\varphi=\pi}> \Delta\textrm{Nu}|_{\varphi=-i\infty}> \Delta\textrm{Nu}|_{\varphi=0} $$
In Fig. \ref {fg5t}, we have depicted the variation of $ \textrm {Nu}(\tau) $ with time $\tau$ for different frequencies of the out-phase ($ \varphi = \pi $) modulation of $ \omega_T = 2,5,10,30 $ and for a fixed Rossby number $ \textrm {Ro} = 2 $ and amplitude $ \delta_1 = 0.5 $.
Here it can be seen that an increase in the modulation frequency $ \omega_T $ leads to suppression of heat transfer, i.e. variations in the number $ \textrm {Nu} $ are reduced:
$$\Delta \textrm{Nu}|_{\omega_T=2}>\Delta \textrm{Nu}|_{\omega_T=5}>\Delta \textrm{Nu}|_{\omega_T=10}>\Delta \textrm{Nu}|_{\omega_T=30}$$
A similar phenomenon is observed for nonuniformly rotation with other profiles.
In Fig. \ref{fg6t}, we have presented the unmodulated  ($\delta_1=0$) result of Eq. (\ref{eq44n}) and compared it with the present results of modulated case for different amplitudes $ \delta_1 = 0.5, 0.3, 0.1 $. 
It can be seen from these graphs that modulation of the temperature boundaries of the layer leads to a periodic change in heat flow, i.e. the value of the number $\textrm{Nu}$ changes periodically in time $\tau$ and increases with increasing amplitude $\delta_1$.

\subsection{Gravity field modulation}

We now turn to the next method of parametric action on stationary magnetoconvection in a nonuniformly rotating electrically conductive medium.
Let  a fluid layer carries out vertical harmonic oscillations with a frequency $ \widetilde {\omega}_g $ and a small   amplitude $ \epsilon^2 \xi $.
Then, in the equations of motion written in the reference frame associated with the fluid layer, the acceleration of gravity $ {\bf g} $ should be replaced by $ {\bf{g}}_0 (1+\epsilon^2 \delta_4 \cos (\widetilde {\omega}_g t)) $ (see formula (\ref{eq3n})).
In the equations (\ref{eq29n}) we set $ \delta_1 = 0 $ and $ f_R = f_m = 1 $.
As a result, the equations of the asymptotic expansion in the third order take the following form:
\begin{equation} \label{eq46n} \widehat{L}M_3=N_3, \end{equation}
\begin{figure}
  \centering
	\includegraphics[width=15 cm,  height=20 cm]{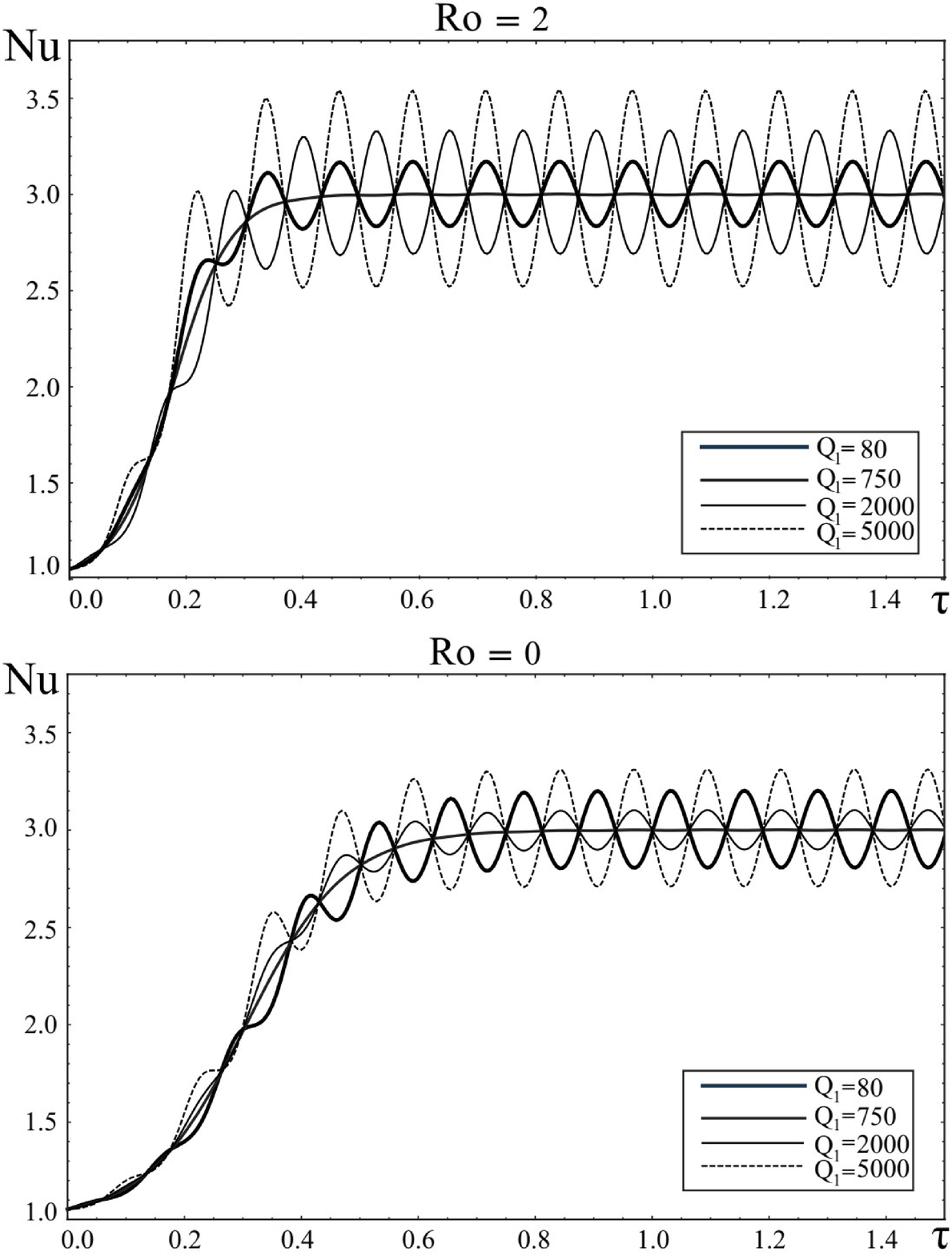} \\
	\caption{Variations of the Nusselt number $ \textrm {Nu} $ depending on the Chandrasekhar number $ \textrm {Q}_1 = (80,750,2000,5000) $ for the amplitude parameters of the oscillating magnetic field $ \delta_3 = 0.5 $, frequency $ \omega_B = 50 $ with positive Rossby numbers $ \textrm {Ro} = (2, 0) $.}\label{fg6m}
	\end{figure}
	\begin{figure}
  \centering
	\includegraphics[width=15 cm,  height=20 cm]{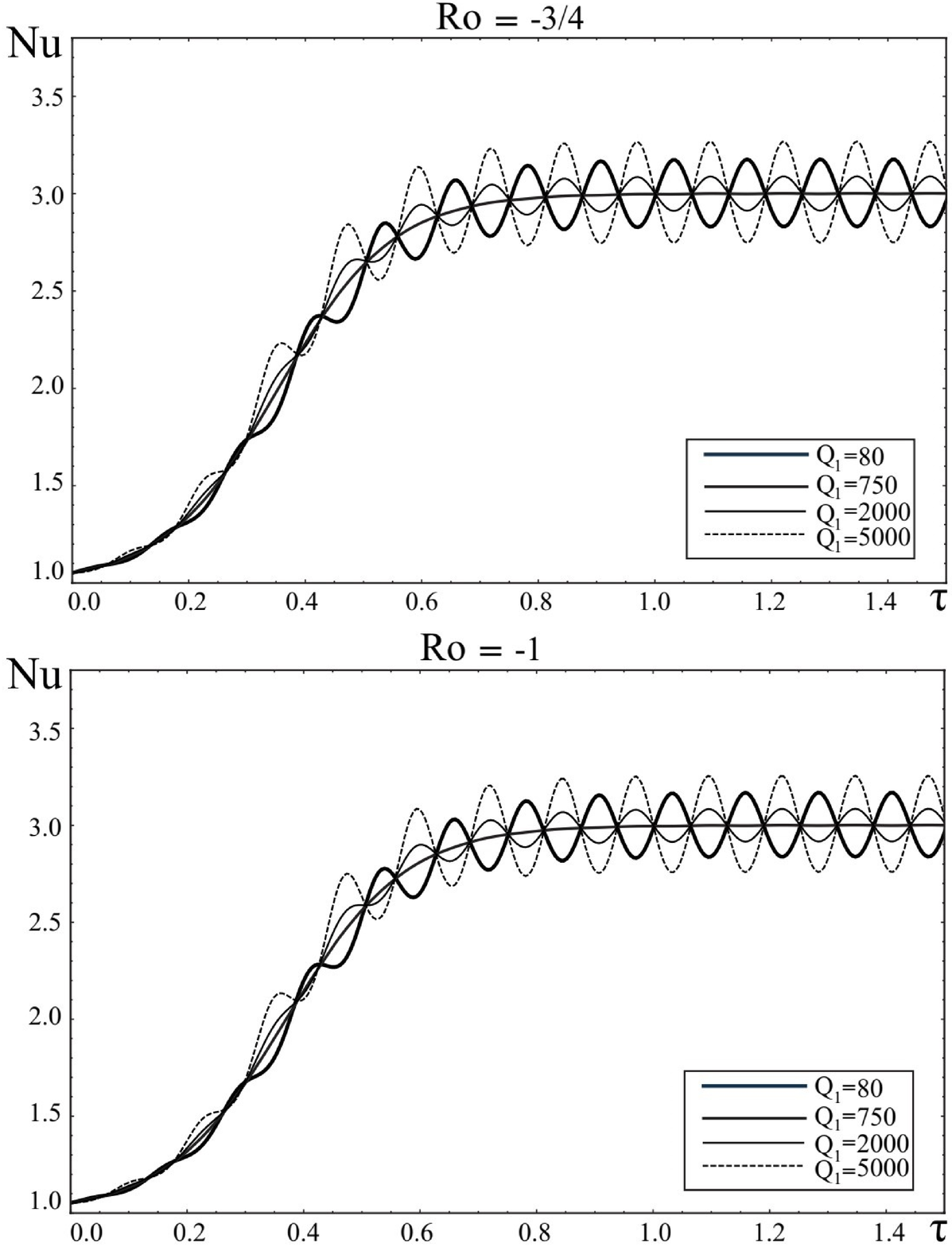} \\
	\caption{Variations of the Nusselt number $ \textrm {Nu} $ depending on the Chandrasekhar number $ \textrm {Q}_1 = (80,750,2000,5000) $ for the amplitude parameters of the oscillating magnetic field $ \delta_3 = 0.5 $, frequency $ \omega_B = 50 $ with negative Rossby numbers $ \textrm {Ro} = (- 3/4, -1) $.}\label{fg6mr}
	\end{figure}
where $M_3=\begin{bmatrix} \psi _3 \\ v_3 \\ \phi _3 \\ \widetilde v_3  \\ \theta _3 \\ \end{bmatrix}$, $N_3=\begin{bmatrix} N_{31} \\ N_{32} \\ N_{33} \\ N_{34}  \\ N_{35} \\ \end{bmatrix}$; 
\[N_{31}=-\frac{\partial}{\partial\tau}\nabla^2\psi_1+\textrm{Ra}_2 \frac{\partial\theta_1}{\partial x}+\textrm{Ra}_c\delta_4\cos(\omega_g \tau)\frac{\partial\theta_1}{\partial x}=\]
\[=\left(a^2\frac{\partial A(\tau)}{\partial \tau}-\textrm{Ra}_2 \frac{k_c^2 A(\tau)}{a^2}-\textrm{Ra}_c\delta_4\cos(\omega_g \tau)\frac{k_c^2 A(\tau)}{a^2}\right)\sin k_c x \sin \pi z,   \]
\[N_{32}=-\frac{\partial v_1}{\partial\tau}=-\frac{\pi \sqrt{\textrm{Ta}}}{a^2}\cdot\frac{(1+\textrm{Ro})a^4+\pi^2 \textrm{QPmRo}}{a^4+\pi^2 \textrm{Q}}\cdot\frac{\partial A(\tau)}{\partial \tau}\sin k_c x \cos \pi z, \]
\[N_{33}=-\frac{\partial\phi_1}{\partial\tau}=-\frac{\pi \textrm{Pm}}{a^2 \Pr}\cdot\frac{\partial A(\tau)}{\partial \tau}\sin k_c x \cos \pi z    \]
\[N_{34}=-\frac{\partial \widetilde v_1}{\partial\tau}= \frac{\pi^2\sqrt{\textrm{Ta}}(1+\textrm{Ro}(1-\textrm{Pm}))\textrm{Pm}}{\Pr(a^4+\pi^2\textrm{Q})}\cdot\frac{\partial A(\tau)}{\partial \tau}\sin k_c x \sin \pi z , \]
\[N_{35}=-\frac{\partial\theta_1}{\partial\tau}-{\Pr}^{-1}\left[\frac{\partial\psi_1}{\partial x}\frac{\partial\theta_2}{\partial z}-\frac{\partial\theta_2}{\partial x}\frac{\partial\psi_1}{\partial z} + \frac{\partial\psi_2}{\partial x}\frac{\partial\theta_1}{\partial z}-\frac{\partial\theta_1}{\partial x}\frac{\partial\psi_2}{\partial z} \right]=\]
\[ =-\frac{k_c}{a^2}\frac{\partial A(\tau)}{\partial \tau}\cos k_c x \sin \pi z+\frac{{\Pr}^{-1}k_c^3}{4a^2} A^3(\tau)\cos k_c x \sin \pi z\cos 2\pi z. \]
We substituted the values of the elements of the matrix $ N_3 $ into the solvability condition (\ref{eq42n}) for the third-order equations $ O (\epsilon^3) $. After completing the integration in (\ref{eq42n}), we have obtained the equation for the evolution of the finite amplitude $ A (\tau) $ in the form of a non-autonomous GL equation:
\begin{equation} \label{eq47n} 
\mathscr A_{1G}\frac{\partial A}{\partial\tau}-\mathscr A_{2G}(\tau) A+\mathscr A_{3G} A^3=0,
 \end{equation}
where the coefficients $ \mathscr A_{1G, 2G, 3G} $ have the form
\begin{equation} \label{eq48n} 
\mathscr A_{1G}=a^2+\frac{k_c^2}{a^4}\Pr\textrm{Ra}_c-\frac{\pi^2}{a^2}\textrm{QPm}-\frac{\pi^2\textrm{Ta}\left((1+\textrm{Ro})a^4+\pi^2\textrm{QPm}(\textrm{RoPm}-1)\right)}{(a^4+\pi^2\textrm{Q})^2}-\frac{\pi^4\textrm{TaRoQ}\textrm{Pm}^2}{a^4(a^4+\pi^2\textrm{Q})}, $$
$$ \mathscr A_{2G}(\tau)=\frac{k_c^2 }{a^2}\textrm{Ra}_c\left(\frac{\textrm{Ra}_2}{\textrm{Ra}_c}+\delta_4\cos(\omega_g \tau)\right),\quad \mathscr A_{3G}=\frac{k_c^4\textrm{Ra}_c}{8a^4}, \quad \omega_g=\frac{\widetilde{\omega}_g}{\epsilon^2}. \end{equation}
Using the numerical solution of the equation (\ref{eq48n}) and the formula (\ref{eq40n}), we have determined  the change in the heat transfer (Nusselt number $ \textrm{Nu} $) from time $ \tau $.
We chose the parameters of the convective medium and the initial amplitude as in the previous section: $$ \textrm {Q} / \pi^2 = \textrm {Q}_1 = 80, \textrm{Ta} / \pi^4 =\textrm{T}_1=10^5, \textrm{Ra}_c / \pi^4 =\textrm{R}_1 = 9500, \textrm{Pm}=1, \textrm{Pr}=10, A_0 = 0.5 .$$
 Fig. \ref{fg3g} shows the time dependence of the Nusselt number $ \textrm {Nu} $ $ \tau $ for different rotation profiles of $\textrm{Ro} = (2, 0, -3/4, -1) $ of the electrically conductive medium (plasma) in an oscillating gravitational field with a frequency of $ \omega_g = 10 $ and an amplitude of $ \delta_4 = 0.3 $.
From Fig. \ref{fg3g} it is clear that the heat transfer in the plasma increases for a nonuniformly rotation with a positive Rossby number $ (\textrm{Ro}> 0) $. This process is well depicted in Fig. \ref{fg3g}.

From Fig. \ref{fg4g} and Fig.\ref{fg4gr}, we can see the effect the variation of $ \textrm {Nu}(\tau) $ with time $\tau$ for different frequencies  modulation frequencies $ \omega_g = 10,25,50,100 $ and for different rotation profiles $ \textrm {Ro} = (2, 0, - 3/4, -1) $ of the electrically conductive medium (plasma). As the frequency increases from 10 to 100, the magnitude of $ \textrm {Nu}(\tau) $ decreases, and the effect of modulation on heat transport diminishes. Therefore, the effect of $ \omega_g$ stabilizes the system: 
$$\Delta \textrm{Nu}|_{\omega_g=10}>\Delta \textrm{Nu}|_{\omega_g=25}>\Delta \textrm{Nu}|_{\omega_g=50}>\Delta \textrm{Nu}|_{\omega_g=100}$$
Now let us compare the heat transfer in the absence of $ \delta_4 = 0 $ and in the presence of modulation $ (\omega_g = 10) $ for different amplitudes $ \delta_4 = 0.5, 0.3, 0.1 $ of the gravitational field. This process is shown in Fig.\ref{fg5g}. Here, the dashed line shows the mode of establishing the final value of $ \textrm {Nu} (\tau) $ for the case $ \delta_4 = 0 $. Here, the dashed line demonstrates the regime of establishing the final value of $ \textrm{Nu} (\tau) $ for the case $ \delta_4 = 0 $. Obviously, the excess of number $ \textrm {Nu}(\tau) $ over the unit is
caused by the convection occurrence. From Fig. \ref{fg5g}  we can see that modulation of the gravitational field leads to a periodic change in heat flow, i.e. the value of the number $ \textrm {Nu} $ changes periodically in time $ \tau $ and increases with increasing amplitude $ \delta_4 $.

In  Fig. \ref{fg6g} and Fig. \ref {fg6gr} we have depicted  the effect of Taylor numbers $ \textrm{T}_1 = \textrm{Ta} / \pi^4 = (10^3,10^4,5 \cdot 10^4,10^5) $  on  Nusselt number  $ \textrm{Nu} $  for fixed values of the modulation frequency of the gravitational field $ \omega_B = 50 $, amplitude $ \delta_4 = 0.5 $ and Rossby numbers $ \textrm{Ro} = (2, 0, -3/4 , -1) $. From Fig. \ref{fg6g} it can be seen that for the Rossby number $ \textrm {Ro} = 2 $, with an increase in the Taylor number $ \textrm {T}_1 $ the heat transfer (the Nusselt number $\textrm{Nu} $) in the system also increases:
$$ \Delta\textrm{Nu}|_{\textrm{T}_1=10^3} < \Delta\textrm{Nu}|_{\textrm{T}_1=10^4}<\Delta\textrm{Nu}|_{\textrm{T}_1=5\cdot 10^4}<\Delta\textrm{Nu}|_{\textrm{T}_1=10^5} $$
 Increasing of the Taylor number for solid-state rotation $ \textrm {Ro}=0 $ has almost no effect on heat transfer. However, for negative rotation profiles $ \textrm {Ro} = -3/4 $ and $ \textrm {Ro} =-1 $ (see Fig. \ref{fg6gr}) with increasing Taylor number $ \textrm{T}_1 $ heat transfer in the system decreases:
$$ \Delta\textrm{Nu}|_{\textrm{T}_1=10^3} > \Delta\textrm{Nu}|_{\textrm{T}_1=10^4}>\Delta\textrm{Nu}|_{\textrm{T}_1=5\cdot 10^4}>\Delta\textrm{Nu}|_{\textrm{T}_1=10^5} $$

\subsection{Modulation of an external magnetic field}

In this section, we have studied the stationary regime of nonlinear magnetoconvection under the influence of a time-dependent magnetic field. We represent the magnetic field as the sum of the constant (stationary) and oscillating parts. The oscillating part has the second order $ \epsilon^2 \cdot \delta_3 $ with respect to the expansion parameter (the supercriticality parameter of the Rayleigh number $ \epsilon $) (see formula (\ref{eq3n})). Assuming in the equations (\ref {eq29n}) $ \delta_1 = 0 $ and $ f_R = f_g = 1 $, the equations of the asymptotic expansion in the third order in $ \epsilon $ take the following  form:
\begin{equation} \label{eq49n} \widehat{L}M_3=N_3, \end{equation}
where $M_3=\begin{bmatrix} \psi _3 \\ v_3 \\ \phi _3 \\ \widetilde v_3  \\ \theta _3 \\ \end{bmatrix}$, $N_3=\begin{bmatrix} N_{31} \\ N_{32} \\ N_{33} \\ N_{34}  \\ N_{35} \\ \end{bmatrix}$; 
\[N_{31}=-\frac{\partial}{\partial\tau}\nabla^2\psi_1+\textrm{Ra}_2 \frac{\partial\theta_1}{\partial x}+\delta_3\cos(\omega_B\tau)\widetilde{\textrm{Q}}\frac{\partial }{\partial z}\nabla^2\phi_1=\]
\begin{figure}
  \centering
	\includegraphics[width=12 cm, height= 8 cm]{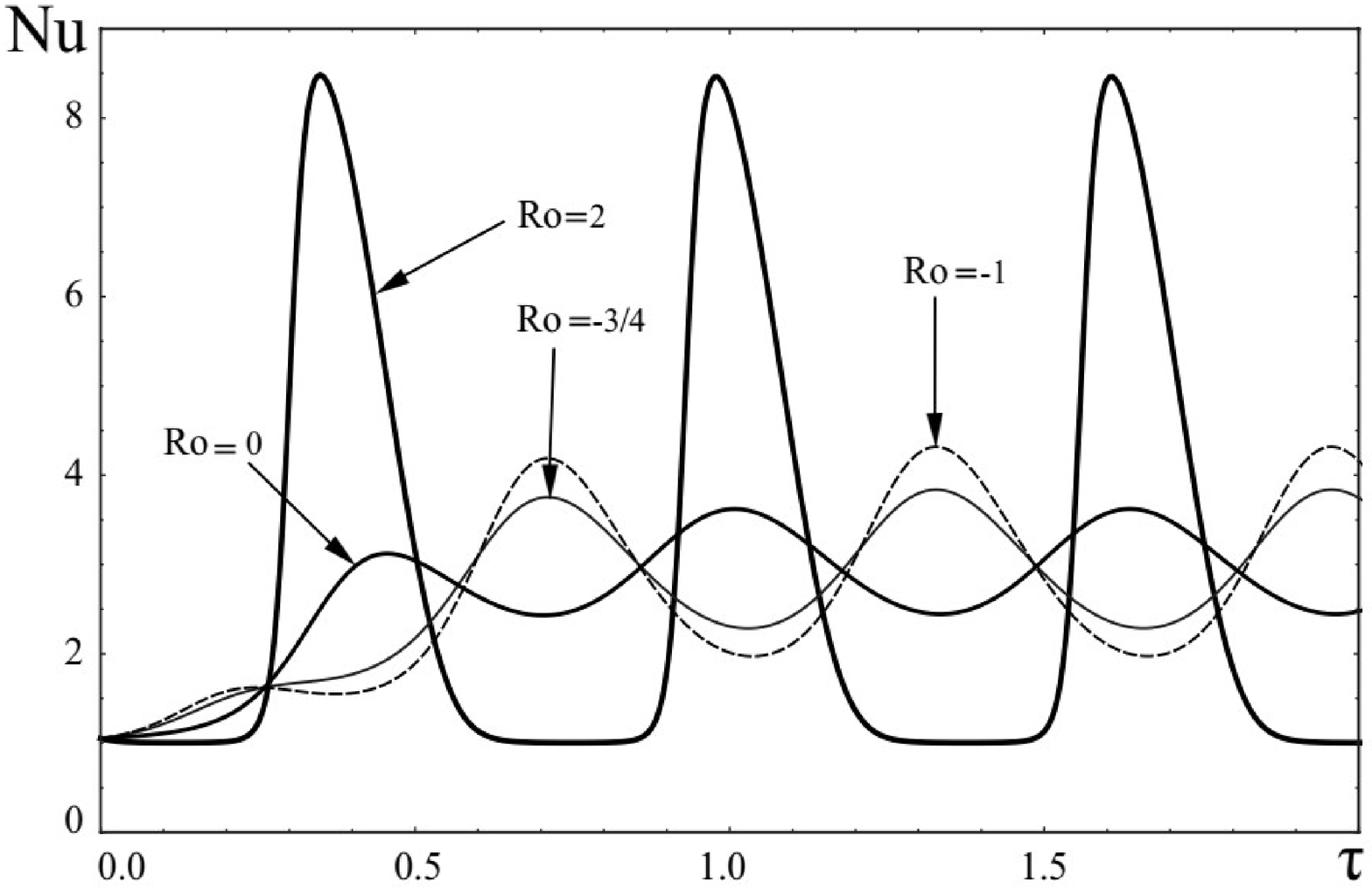}   \\
	\caption{Dependency of the Nusselt number $ \textrm {Nu} $ on the time $ \tau $ for Rossby numbers $ \textrm {Ro} = (2, 0, -3/4, -1) $ when modulating rotation with a frequency of $ \omega_R = 10 $ and amplitude $ \delta = 0.3 $. }\label{fg3r}
	\end{figure}
\[=\left(a^2\frac{\partial A(\tau)}{\partial \tau}-\textrm{Ra}_2 \frac{k_c^2 A(\tau)}{a^2}+\pi^2\textrm{Q}\delta_3\cos(\omega_B\tau)\right)\sin k_c x \sin \pi z, \;  \widetilde{\textrm{Q}}=\Pr\textrm{Pm}^{-1}\textrm{Q},  \]
\[N_{32}=-\frac{\partial v_1}{\partial\tau}+\delta_3\cos(\omega_B\tau)\widetilde{\textrm{Q}}\frac{\partial \widetilde{v}_1}{\partial z} =\]
\[=-\frac{\pi\sqrt{\textrm{Ta}}}{a^2(a^4+\pi^2 \textrm{Q})}\left[((1+\textrm{Ro})a^4+\pi^2\textrm{QPmRo})\frac{\partial A(\tau)}{\partial \tau}+
\right.$$
$$\left.+\pi^2a^2\textrm{Q}(1+\textrm{Ro}(1-\textrm{Pm}))\delta_3\cos(\omega_B\tau)A(\tau)\right]\sin k_c x \cos \pi z, \]
\[N_{33}=-\frac{\partial\phi_1}{\partial\tau}+\textrm{Pr}^{-1}\delta_3\cos(\omega_B\tau) \frac{\partial \psi_1}{\partial z}=\]
\[=\left[-\frac{\pi\textrm{Pm}}{a^2\Pr}\frac{\partial A(\tau)}{\partial \tau}+\delta_3\cos(\omega_B\tau)\frac{\pi A(\tau)}{\Pr} \right]\sin k_c x \cos \pi z, \]
\[N_{34}=-\frac{\partial \widetilde v_1}{\partial\tau}+\textrm{Pr}^{-1}\delta_3\cos(\omega_B\tau) \frac{\partial v_1}{\partial z}=\]
\[= \frac{\pi^2\sqrt{\textrm{Ta}}}{a^2\Pr(a^4+\pi^2\textrm{Q})}\left[a^2\textrm{Pm}(1+\textrm{Ro}(1-\textrm{Pm}))\frac{\partial A(\tau)}{\partial \tau}-\right.$$
$$\left.-\delta_3\cos(\omega_B\tau)((1+\textrm{Ro})a^4+\pi^2\textrm{QPmRo})A(\tau)\right]\sin k_c x \sin \pi z , \]
\[N_{35}=-\frac{\partial\theta_1}{\partial\tau}-{\Pr}^{-1}\left[\frac{\partial\psi_1}{\partial x}\frac{\partial\theta_2}{\partial z}-\frac{\partial\theta_2}{\partial x}\frac{\partial\psi_1}{\partial z} + \frac{\partial\psi_2}{\partial x}\frac{\partial\theta_1}{\partial z}-\frac{\partial\theta_1}{\partial x}\frac{\partial\psi_2}{\partial z} \right]=\]
\[ =-\frac{k_c}{a^2}\frac{\partial A(\tau)}{\partial \tau}\cos k_c x \sin \pi z+\frac{{\Pr}^{-1}k_c^3}{4a^2} A^3(\tau)\cos k_c x \sin \pi z\cos 2\pi z. \]
\begin{figure}
  \centering
	\includegraphics[width=15 cm, height=20 cm]{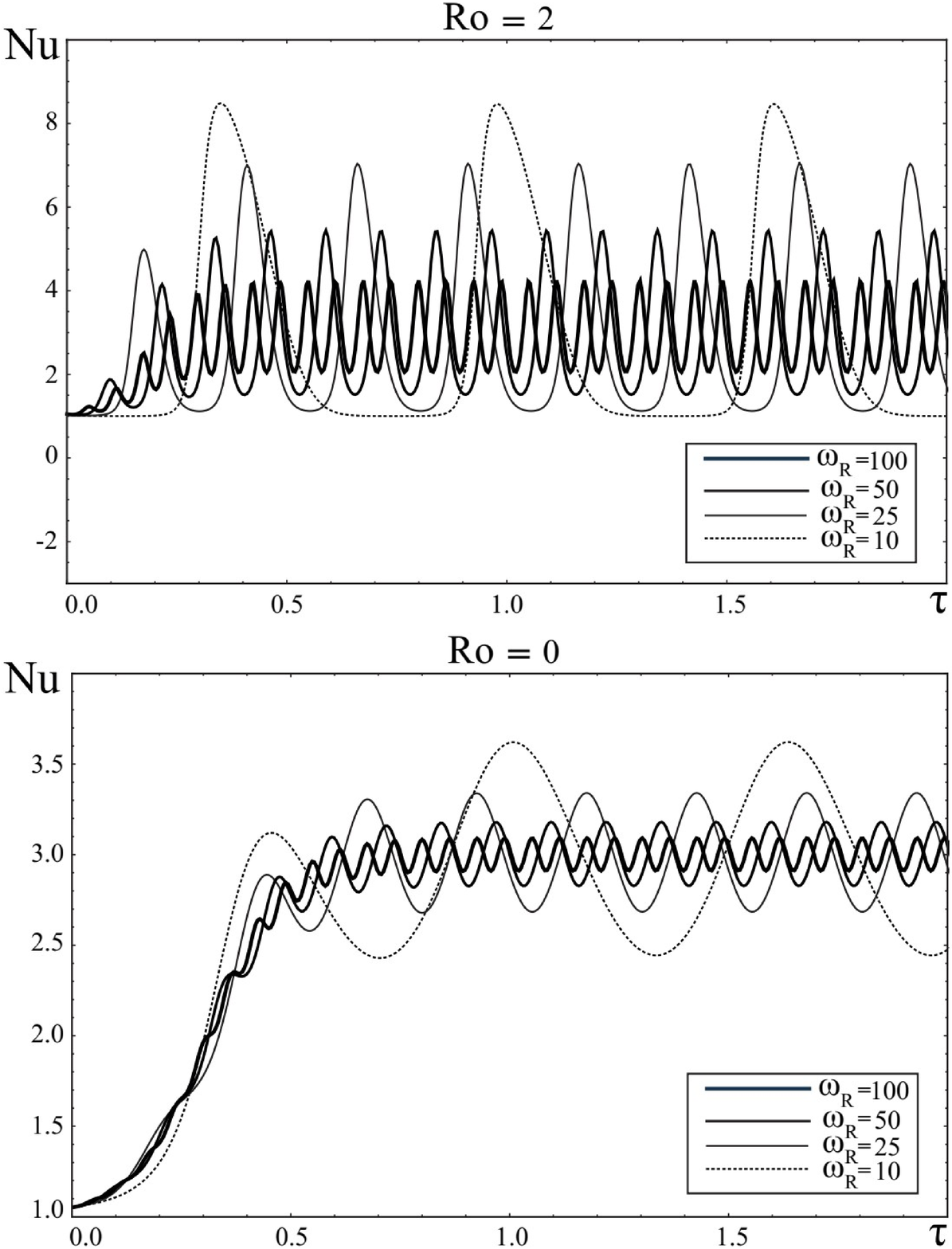}   \\
	\caption{Dependency of the Nusselt number $ \textrm{Nu} $ on time $ \ tau $ for positive Rossby numbers $ \textrm{Ro} =(2, 0) $ for modulation of rotation with a frequency $ \omega_R = (10, 25, 50, 100 ) $ and the amplitude $ \delta = 0.3 $. }\label{fg4r}
	\end{figure}
	\begin{figure}
  \centering
	\includegraphics[width=15 cm, height=20 cm]{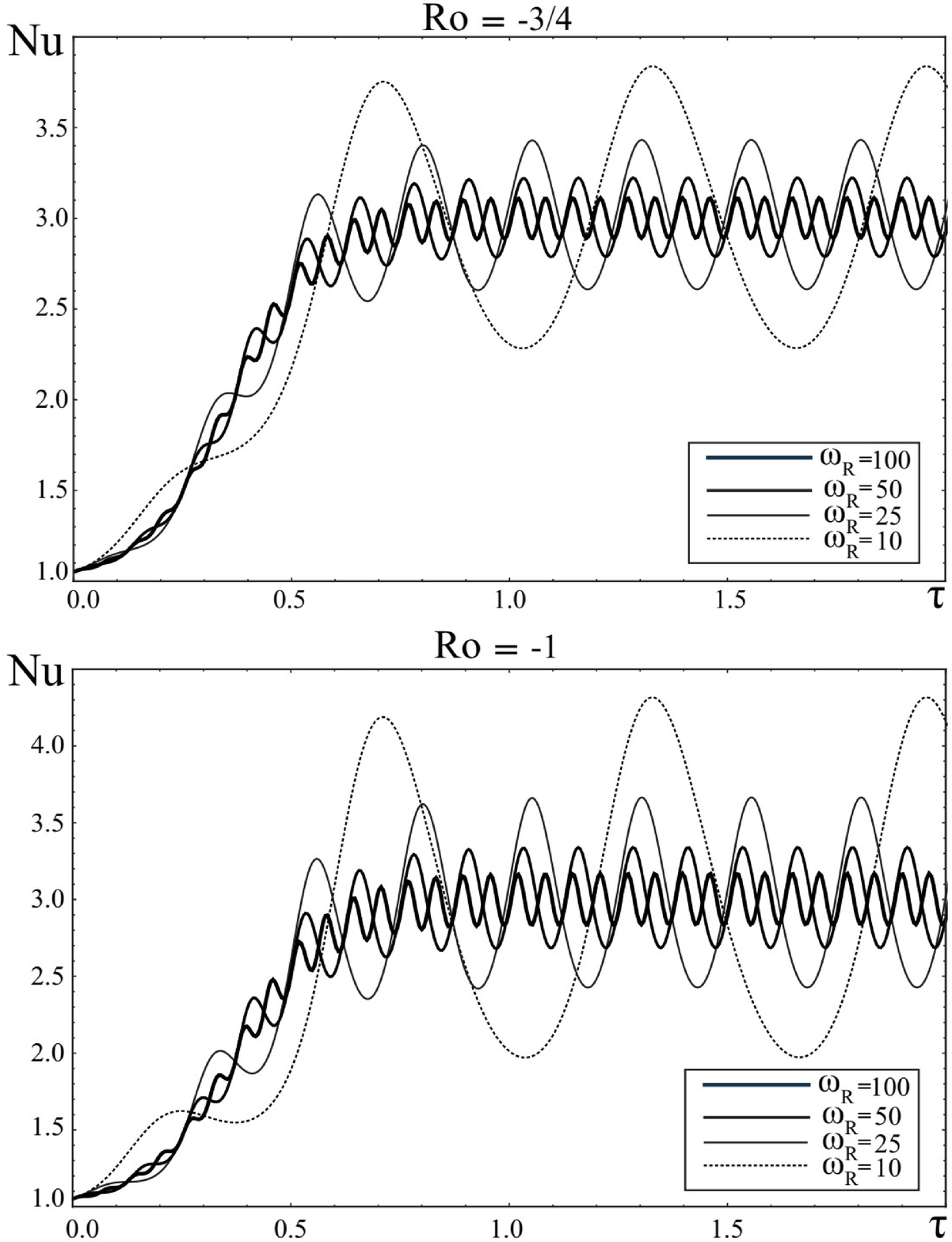}   \\
	\caption{Dependency of the Nusselt number $ \textrm {Nu} $ on time $ \tau $ for negative Rossby numbers $ \textrm {Ro} = (- 3/4, -1) $ when modulating rotation with a frequency of $ \omega_R = (10, 25 , 50, 100) $ and the amplitude $ \delta = 0.3 $.}\label{fg4rt}
	\end{figure}
Using the solvability condition (\ref {eq42n}) and the values of the elements of the matrix $ N_3 $, we can obtain the equation for the evolution of the finite amplitude $ A (\tau) $ in the form of a non-autonomous GL equation:
\begin{equation} \label{eq50n} 
\mathscr A_{1M}\frac{\partial A}{\partial\tau}-\mathscr A_{2M}(\tau) A+\mathscr A_{3M} A^3=0,
 \end{equation}
where the coefficients $ \mathscr A_{1M, 2M, 3M} $ have the form
\begin{equation} \label{eq51n} 
\mathscr A_{1M}=a^2+\frac{k_c^2}{a^4}\Pr\textrm{Ra}_c-\frac{\pi^2}{a^2}\textrm{QPm}-\frac{\pi^2\textrm{Ta}\left((1+\textrm{Ro})a^4+\pi^2\textrm{QPm}(\textrm{RoPm}-1)\right)}{(a^4+\pi^2\textrm{Q})^2}-\frac{\pi^4\textrm{TaRoQ}\textrm{Pm}^2}{a^4(a^4+\pi^2\textrm{Q})}, $$
$$ \mathscr A_{2M}(\tau)=\frac{k_c^2 }{a^2}\textrm{Ra}_2-2\textrm{Q}\pi^2\delta_3\cos(\omega_B\tau)+$$
$$+\frac{\pi^4\textrm{QTa}}{a^2(a^4+\pi^2\textrm{Q})}\left[\frac{1}{a^4+\pi^2\textrm{Q}}\left(2(1+\textrm{Ro})a^4+\textrm{RoPm}(\pi^2\textrm{Q}-a^4)\right)-\textrm{PmRo}\right]\delta_3\cos(\omega_B\tau), $$
$$\omega_B=\frac{\widetilde{\omega}_B}{\epsilon^2},\quad  A_{3M}=\frac{k_c^4\textrm{Ra}_c}{8a^4}.
\end{equation}
In the limiting case, when there is no rotation ($ \textrm {Ta} = 0, \textrm {Ro} = 0 $), the equation (\ref{eq50n}) coincides with the result of \cite {33s}. 

Then, let us perform  a numerical analysis of the equation (\ref{eq50n}) for different rotation profiles $ \textrm {Ro} = (2, 0, -3/4, -1) $ for fixed convection parameters $ \textrm {Q} / \pi^2 = \textrm{Q} _1 = $ 80, $ \textrm {Ta} / \pi^4 = \textrm {T}_1 = 10^5 $, $ \textrm{Ra}_c / \pi^4 = \textrm{R} _1 = 9500 $, $ \textrm {Pm} = 1 $, $ \textrm {Pr} = 10 $, $ A_0 = 0.5 $ and magnetic field modulation $ \omega_B = 10 $, $ \delta_3 = 0.3 $. From  Fig. \ref {fg3m} it can be seen that for nonuniformly rotation with a positive Rossby number $ (\textrm {Ro}> 0) $, the variations of the heat flux in the plasma increase:
$$ \Delta\textrm{Nu}|_{\textrm{Ro}=-1}\leq \Delta\textrm{Nu}|_{\textrm{Ro}=-3/4}<\Delta\textrm{Nu}|_{\textrm{Ro}=0}<\Delta\textrm{Nu}|_{\textrm{Ro}=2} $$
In Fig. \ref{fg4m} and Fig. \ref{fg4mr}, the Nusselt number  $ \textrm {Nu} (\tau) $ with respect to time  $\tau$ has been plotted for different modulation frequencies $ \omega_B = 10,25,50,100 $ and for different rotation profiles $ \textrm {Ro} = (2, 0, - 3/4, -1) $ of the electrically conductive medium (plasma). Here we see that an increase of the modulation frequency $ \omega_B $  leads to suppression of heat transfer  for  the different  Rossby numbers $ \textrm {Ro} $, i.e. variations in the number $ \textrm {Nu} $ are reduced:
$$\Delta \textrm{Nu}|_{\omega_g=10}>\Delta \textrm{Nu}|_{\omega_g=25}>\Delta \textrm{Nu}|_{\omega_g=50}>\Delta \textrm{Nu}|_{\omega_g=100}$$
 Fig. \ref {fg5m} shows the dependency of the heat transfer value $ \textrm {Nu} $ on $ \tau $ in the absence of $ \delta_3 = 0 $ and in the presence of modulation of the magnetic field $ (\omega_B = 10) $ for different amplitudes $ \delta_3 = 0.5, 0.3, 0.1 $.
 The dashed line shows the regime of establishing the final value of $ \textrm {Nu} (\tau) $ for the case $ \delta_3 = 0 $. In Fig. \ref {fg5m} we can see that the modulation of the magnetic field leads to a periodic change of the heat flux, which increases with increasing amplitude $ \delta_3 $.

In Fig. \ref{fg6m} -Fig. \ref {fg6mr} we have depicted the effect of the external magnetic field (Chandrasekhar number $ \textrm {Q}_1 $) on the heat transfer  with an oscillating magnetic field of frequency $ \omega_B = 10 $ and amplitude $ \delta_3 = 0.3 $ for different profiles of nonuniform rotation of $ (\textrm {Ro} = 2, 0, -3/4, -1) $.  Increasing  the external magnetic field, i.e. Chandrasekhar numbers from $ \textrm {Q}_1 = $ 80 to $ \textrm {Q}_1 = $ 750  for $ \textrm {Ro} = 2 $ leads to a decrease of the heat flow, which is almost stabilized at a certain level.
\begin{figure}
  \centering
	\includegraphics[width=12 cm, height=8 cm]{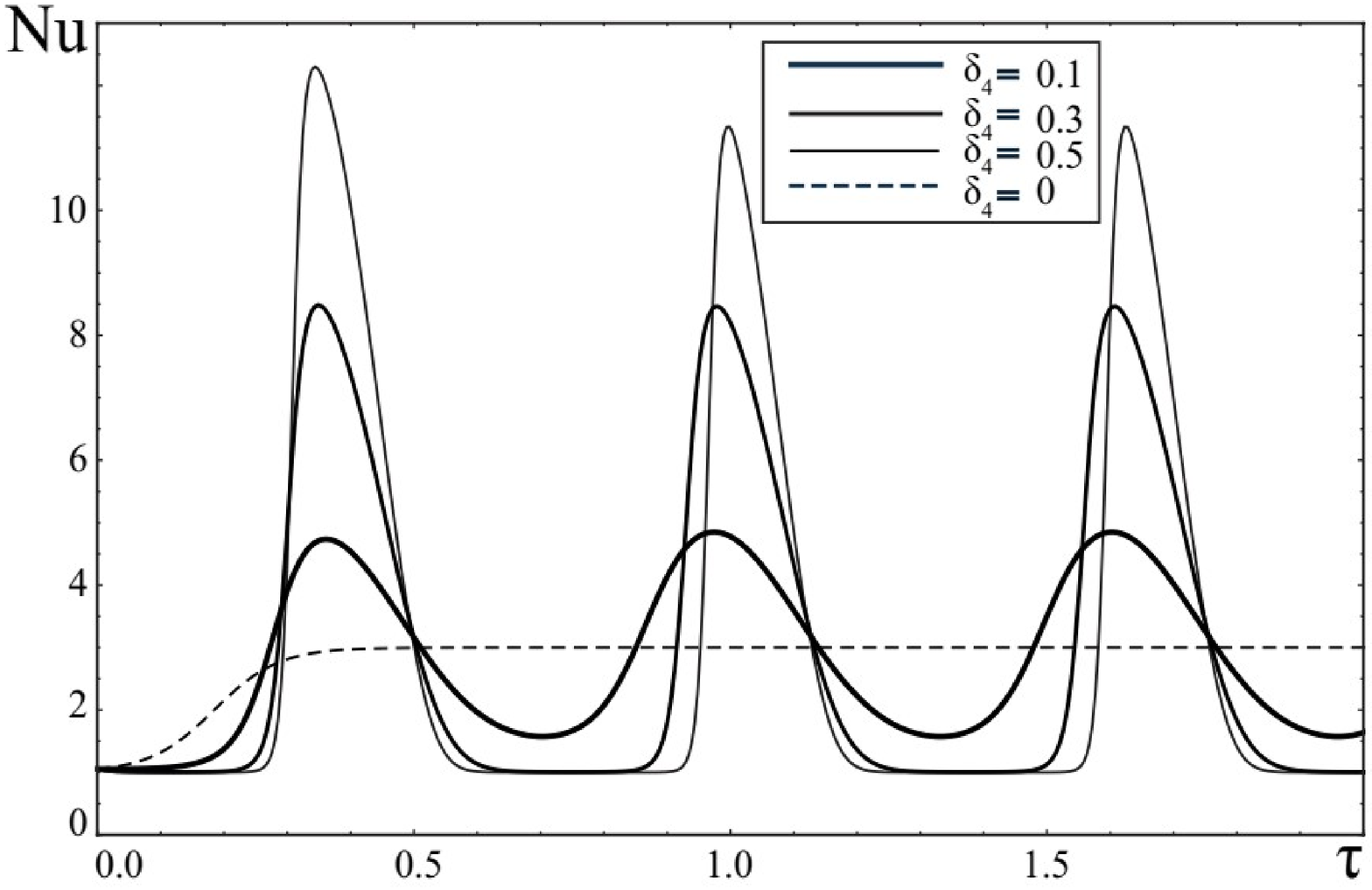}   \\
	\caption{Variations of the Nusselt number $ \textrm {Nu} $ depending on the amplitude of the modulation of rotation $ \delta_4 = (0, 0.1, 0.3, 0.5) $ for the Rossby number $ \textrm {Ro} = 2 $ and the frequency $ \omega_R = 10 $. }\label{fg5r}
	\end{figure}
The Nusselt number Nu increases with the increasing the value of the magnetic field 	
$$ \Delta\textrm{Nu}|_{\textrm{Q}_1=80}< \Delta\textrm{Nu}|_{\textrm{Q}_1=2000}<\Delta\textrm{Nu}|_{\textrm{Q}_1=5000}$$	
Similar picture  can be obtained for other rotation profiles $ \textrm {Ro} = (0, -3/4, -1) $: 
$$ \Delta \textrm {Nu}|_ {\textrm {Q}_1 = 80} <\Delta \textrm {Nu} |_ {\textrm {Q}_1 = 5000} .$$

\subsection{Rotational speed modulation}

One of the important ways of parametric influence on convective instability is the modulation of the angular velocity of rotation.
The first experiments on rotation modulation were carried out in \cite {42s}. It was also found there the onset of instability in Couette flow can be inhibited by modulating the rate of rotation of the inner cylinder. Later, numerical \cite {35s} and laboratory experiments \cite {43s} were carried out for the effect of rotation modulation on Rayleigh-Benard convection.  In \cite{35s} it is shown that the modulation of the rotation of $ \Omega (t) = \Omega_0 + \Delta \Omega \sin (\omega t) $ causes a periodic change in the Coriolis force, and therefore has an effect on the onset of convection.
In \cite{43s}, the experiments were carried out with a periodic dependency of the angular velocity $ \Omega (t) = \Omega_0 (1 \beta \cos (\omega t)) $ ($ \Omega_0 = 0.104 $ $ rad / s $, $ \beta = $ 0.212) for modeling large-scale atmospheric flow dynamics in order to create a longer-term weather forecast.

Here we also assume that the oscillating part is of the second order of smallness $ \epsilon^2 \cdot \delta_3 $ in the expansion parameter (the supercriticality of the Rayleigh number $ \epsilon $) (see the formula (\ref{eq2n})). Assuming in the equations (\ref{eq29n}) $ \delta_1 = 0 $ and $ f_m = f_g = 1 $, the equations of the asymptotic expansion in the third order ($ \epsilon^3 $) take the form
\begin{equation} \label{eq52n} \widehat{L}M_3=N_3, \end{equation}
where $M_3=\begin{bmatrix} \psi _3 \\ v_3 \\ \phi _3 \\ \widetilde v_3  \\ \theta _3 \\ \end{bmatrix}$, $N_3=\begin{bmatrix} N_{31} \\ N_{32} \\ N_{33} \\ N_{34}  \\ N_{35} \\ \end{bmatrix}$; 
\[N_{31}=-\frac{\partial}{\partial\tau}\nabla^2\psi_1+\textrm{Ra}_2 \frac{\partial\theta_1}{\partial x}-\delta_2\cos(\omega_R\tau)\,\sqrt{\textrm{Ta}}\frac{\partial v_1}{\partial z} =\]
\begin{figure}
  \centering
	\includegraphics[width=15 cm,  height=20 cm]{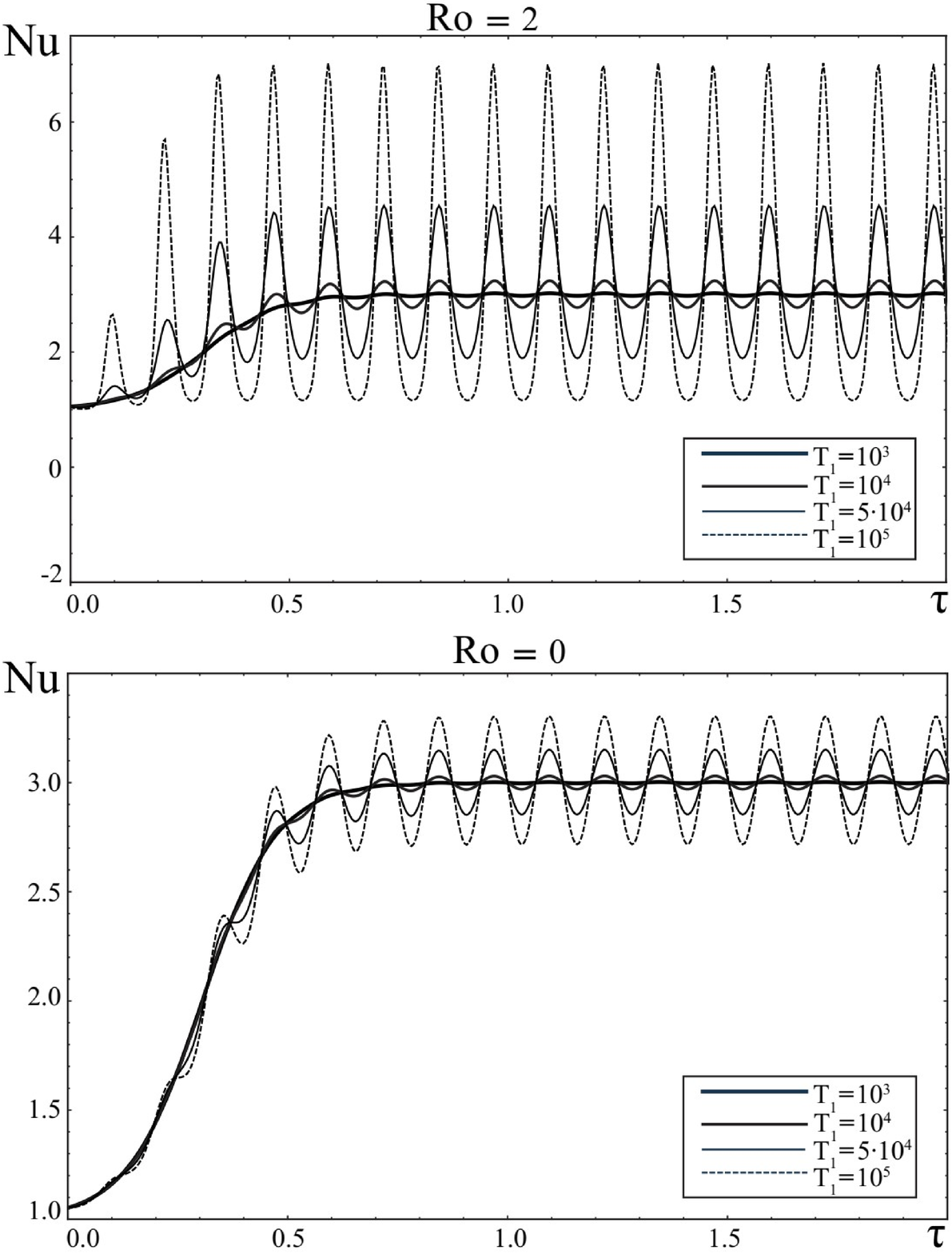} \\
	\caption{Variations of the Nusselt number $ \textrm{Nu} $ depending on the Taylor number $ \textrm{T}_1 = (10^3, 10^4, 5 \cdot 10^4, 10^5) $ for rotation modulation parameters ($ \delta_2 = 0.5 $, $ \omega_R = 50 $) with positive Rossby numbers $ \textrm {Ro} = (2, 0) $.}\label{fg6r}
	\end{figure}
	\begin{figure}
  \centering
	\includegraphics[width=15 cm,  height=20 cm]{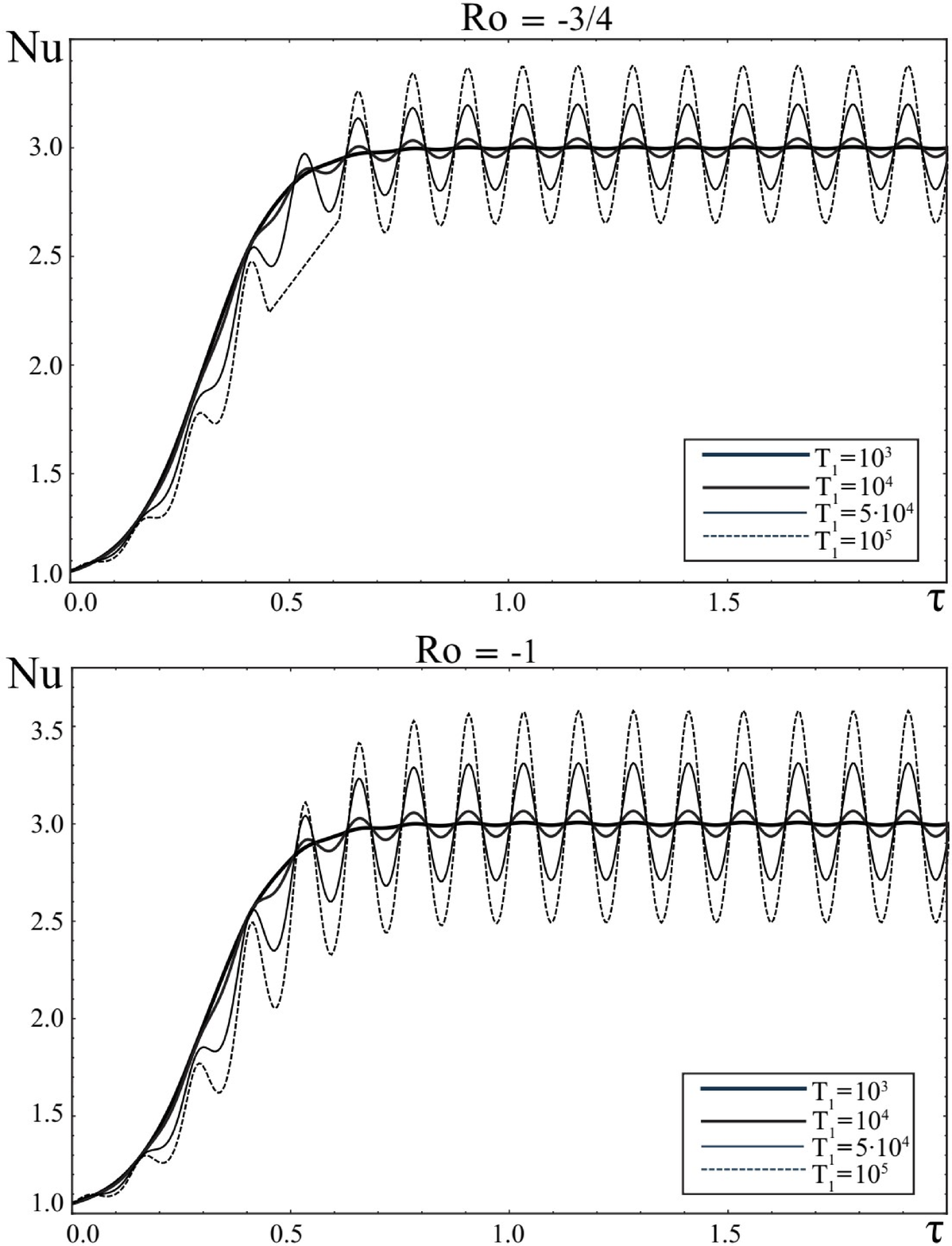} \\
	\caption{Variations of the Nusselt number $ \textrm {Nu} $ depending on the Taylor number $ \textrm {T} _1 = (10^3, 10^4, 5 \cdot 10^4, 10^5) $ for rotation modulation parameters ($ \delta_2 = 0.5 $, $ \omega_R = 50 $) for negative Rossby numbers $ \textrm {Ro} = (- 3/4, -1) $.}\label{fg6rt}
	\end{figure}
\[=\left(a^2\frac{\partial A(\tau)}{\partial \tau}-\textrm{Ra}_2 \frac{k_c^2 A(\tau)}{a^2}+\delta_2\cos(\omega_R\tau)\cdot\frac{\pi^2\textrm{Ta}((1+\textrm{Ro})a^4+\pi^2\textrm{QPmRo})A(\tau)}{a^2(a^4+\pi^2 \textrm{Q})}\right)\sin k_c x \sin \pi z,\]
\[N_{32}=-\frac{\partial v_1}{\partial\tau}+\delta_2\cos(\omega_R\tau)\sqrt{\textrm{Ta}}(1+\textrm{Ro})\frac{\partial\psi_1}{\partial z}= \] 
\[=\left[-\frac{\pi\sqrt{\textrm{Ta}}}{a^2(a^4+\pi^2 \textrm{Q})}\left((1+\textrm{Ro})a^4+\pi^2\textrm{QPmRo}\right)\frac{\partial A(\tau)}{\partial \tau}+\pi\sqrt{\textrm{Ta}}(1+\textrm{Ro})\delta_2\cos(\omega_R\tau)A(\tau)\right]\sin k_c x \cos \pi z, \]
\[N_{33}=-\frac{\partial\phi_1}{\partial\tau}=-\frac{\pi \textrm{Pm}}{a^2 \Pr}\cdot\frac{\partial A(\tau)}{\partial \tau}\sin k_c x \cos \pi z, \]
\[N_{34}=-\frac{\partial \widetilde v_1}{\partial\tau}-\delta_2\cos(\omega_R\tau)\textrm{Ro}\sqrt{\textrm{Ta}}\frac{\partial\phi_1}{\partial z}=\]
\[= \left[\frac{\pi^2\sqrt{\textrm{Ta}}(1+\textrm{Ro}(1-\textrm{Pm}))\textrm{Pm}}{\Pr(a^4+\pi^2\textrm{Q})}\frac{\partial A(\tau)}{\partial \tau}+\delta_2\cos(\omega_R\tau)\textrm{Ro}\sqrt{\textrm{Ta}}\frac{\pi^2\textrm{Pm}}{a^2\Pr}A(\tau)\right]\sin k_c x \sin \pi z,  \]
\[N_{35}=-\frac{\partial\theta_1}{\partial\tau}-{\Pr}^{-1}\left[\frac{\partial\psi_1}{\partial x}\frac{\partial\theta_2}{\partial z}-\frac{\partial\theta_2}{\partial x}\frac{\partial\psi_1}{\partial z} + \frac{\partial\psi_2}{\partial x}\frac{\partial\theta_1}{\partial z}-\frac{\partial\theta_1}{\partial x}\frac{\partial\psi_2}{\partial z} \right]=\]
\[ =-\frac{k_c}{a^2}\frac{\partial A(\tau)}{\partial \tau}\cos k_c x \sin \pi z+\frac{{\Pr}^{-1}k_c^3}{4a^2} A^3(\tau)\cos k_c x \sin \pi z\cos 2\pi z. \]
By substituting the values of the elements of the matrix $ N_3 $ into the solvability condition (\ref{eq42n}) for the third-order equations $ O (\epsilon^3) $, we can obtain the equation for the evolution of the finite amplitude $ A (\tau) $ in the form of a non-autonomous GL equation:
\begin{equation} \label{eq53n} 
\mathscr A_{1R}\frac{\partial A}{\partial\tau}-\mathscr A_{2R}(\tau) A+\mathscr A_{3R} A^3=0,
 \end{equation}
where the coefficients $ \mathscr A_{1R, 2R, 3R} $ have the following form
\begin{equation} \label{eq54n} 
\mathscr A_{1R}=a^2+\frac{k_c^2}{a^4}\Pr\textrm{Ra}_c-\frac{\pi^2}{a^2}\textrm{QPm}-\frac{\pi^2\textrm{Ta}\left((1+\textrm{Ro})a^4+\pi^2\textrm{QPm}(\textrm{RoPm}-1)\right)}{(a^4+\pi^2\textrm{Q})^2}-\frac{\pi^4\textrm{TaRoQ}\textrm{Pm}^2}{a^4(a^4+\pi^2\textrm{Q})}, $$
$$ \mathscr A_{2R}(\tau)=\frac{k_c^2 }{a^2}\textrm{Ra}_2-\delta_2\cos(\omega_R\tau)\cdot\frac{2\pi^2\textrm{Ta}((1+\textrm{Ro})a^4+\pi^2\textrm{QRoPm})}{a^2(a^4+\pi^2 \textrm{Q})},\; \omega_R=\frac{\widetilde{\omega}_R}{\epsilon^2}, \; \mathscr A_{3R}=\frac{k_c^4\textrm{Ra}_c}{8a^4}. \end{equation}
Here we carry out a numerical analysis of the equation (\ref{eq50n}) for different rotation profiles $ (\textrm {Ro} = 2, 0, -3/4, -1) $ and we present the results of the dependence of the heat transfer (Nusselt number $ \textrm {Nu} $) from the time $ \tau $ (see Fig. \ref{fg3r}).
We consider the convection parameters to be fixed $ \textrm {Q} / \pi^2 = \textrm {Q}_1 = $ 80, $ \textrm{Ta} / \pi^4 = \textrm{T}_1 = 10^5 $, $ \textrm{Ra}_c / \pi^4 = \textrm {R}_1 = 9500 $, $ \textrm {Pm}= 1 $, $ \textrm{Pr} = 10 $, $ A_0 = 0.5 $ and rotation speed modulation has a frequency and an amplitude: $ \omega_R = 10 $ and $ \delta_3 = 0.3 $. 
From Fig. \ref{fg3r} , we can observe that for  nonuniform rotation with a positive Rossby number $ (\textrm {Ro} = 2) $ the variations of heat flow are greater than  for  nonuniform rotation with negative Rossby numbers $ (\textrm {Ro} = -3/4 , -1) $ and $ (\textrm{Ro} = 0) $:
$$ \Delta\textrm{Nu}|_{\textrm{Ro}=0} < \Delta\textrm{Nu}|_{\textrm{Ro}=-3/4}<\Delta\textrm{Nu}|_{\textrm{Ro}=-1}<\Delta\textrm{Nu}|_{\textrm{Ro}=2} $$
In Fig. \ref{fg4r}-Fig. \ref{fg4rt} , we have depicted  that  for higher value of the modulation frequency $ \omega_R $ for different Rossby numbers $ \textrm {Ro} = (2, 0, -3/4, -1) $ leads to suppression of heat transfer, i.e. the magnitude of  $\textrm{Nu}$  decreases:
$$\Delta \textrm{Nu}|_{\omega_R=10}>\Delta \textrm{Nu}|_{\omega_R=25}>\Delta \textrm{Nu}|_{\omega_R=50}>\Delta \textrm{Nu}|_{\omega_R=100}$$
\begin{figure}
  \centering
	\includegraphics[width=15 cm,  height=20 cm]{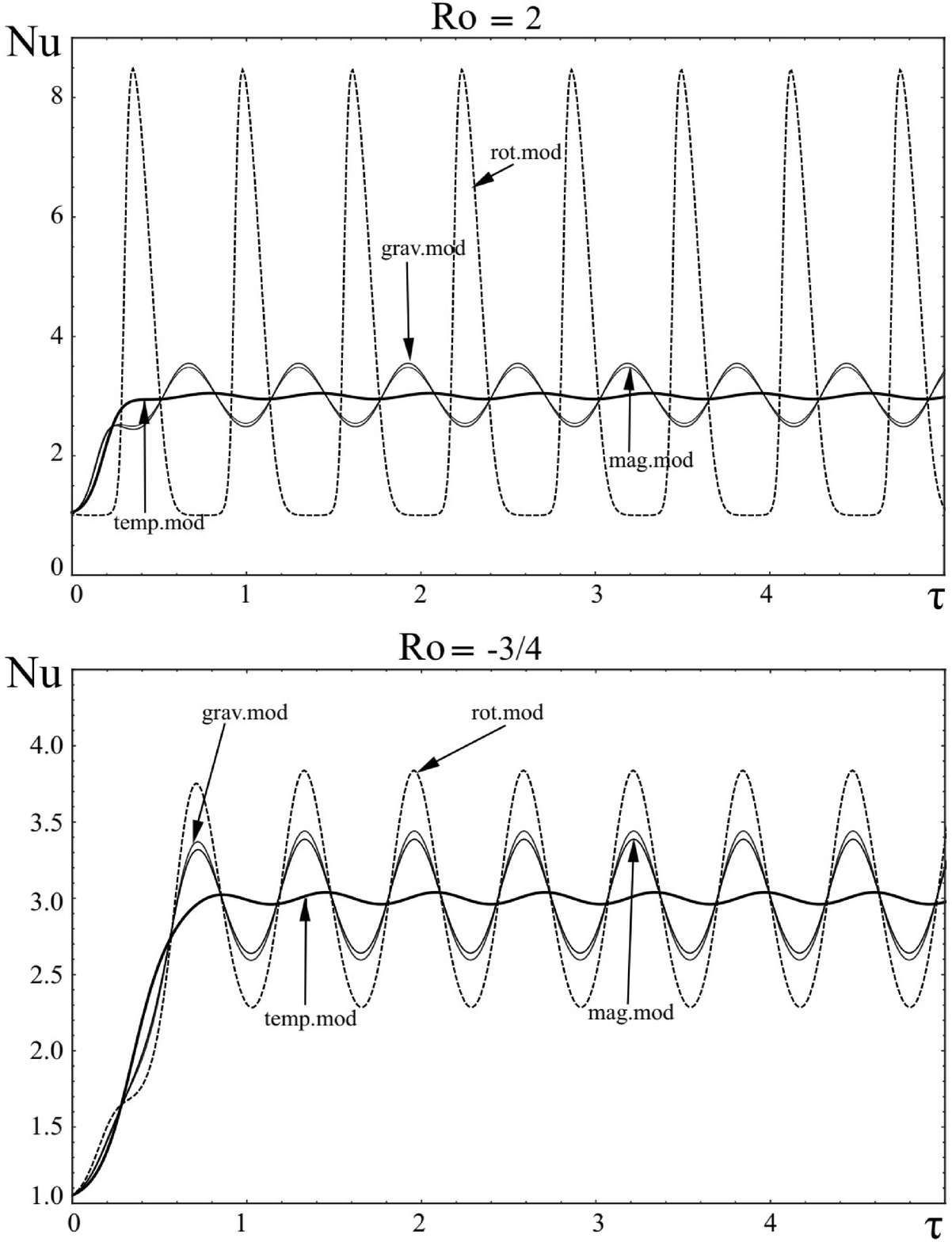} \\
	\caption{Dependency of the Nusselt number $ \textrm {Nu} $ on $ \tau $ for various types of modulation for positive $ \textrm {Ro} = 2 $ and negative $ \textrm {Ro} = - 3/4 $ Rossby numbers.}\label{fg4}
	\end{figure}
In  Fig. \ref{fg5r}, we have defined  the dependency of the heat transfer  $ \textrm {Nu} $ on $ \tau $ in the absence of $ \delta_2 = 0 $ and in the presence of  rotational speed modulation $ (\omega_R = 10) $ for different amplitudes $ \delta_2 = 0.5, 0.3, 0.1 $.
Here, the dashed line depicts the regime of establishing the final value of $ \textrm {Nu} (\tau) $ for the case $ \delta_2 = 0 $. In Fig. \ref{fg5r}, we can see that modulation of rotation leads to a periodic change of the heat flow, which increases with increasing amplitude $ \delta_2 $.

From Fig. \ref{fg6r}  and Fig. \ref {fg6rt}, we can determine the change of heat transfer $ \textrm {Nu} $  for different Taylor numbers $ \textrm {T} _1 = \textrm {Ta} / \pi^4 = (10^3,10^4,5 \cdot 10^4,10^5) $ for a fixed value of the rotation modulation frequency $ \omega_R = 50 $, amplitude $ \delta_2 = 0.5 $ and Rossby numbers $ \textrm {Ro} = (2, 0, -3/4, -1) $.
It can be seen from these graphs that for different Rossby numbers $ \textrm {Ro} = (2, 0, -3/4, -1) $, with an increase in the Taylor number $ \textrm {T}_1 $ heat transfer (Nusselt number $ \textrm{Nu} $) in the system also increases:
$$ \Delta\textrm{Nu}|_{\textrm{T}_1=10^3} < \Delta\textrm{Nu}|_{\textrm{T}_1=10^4}<\Delta\textrm{Nu}|_{\textrm{T}_1=5\cdot 10^4}<\Delta\textrm{Nu}|_{\textrm{T}_1=10^5}. $$

\section{Conclusion}

A weakly nonlinear theory of stationary convection in a nonuniformly  rotating electrically conductive fluid  with a vertical constant magnetic field under the parametric action of: a) temperature modulation of the layer boundaries, b) modulation of gravity, c) modulation of an external magnetic field, d) modulation of the angular velocity of rotation  is developed. We studied the influence of time-periodic modulation on stationary Rayleigh-Benard convection using the perturbation theory method for the small parameter of supercriticality of the Rayleigh number $ \epsilon = \sqrt {(\textrm{Ra}-\textrm{Ra}_c) / \textrm {Ra}_c} $. Furthermore, we considered the amplitudes of the modulated fields to be small, which having a second order $ O (\epsilon^2) $. In the first order $ \epsilon $, the parametric effect does not influence to the development of convection and we obtained the result of the linear theory \cite {11s}.  In the third order $ \epsilon^3 $,  the nonlinear Ginzburg-Landau equation with time-periodic coefficients for four types of modulation are obtained. A numerical analysis of these equations have shown a number of general laws:
\begin{enumerate}
\item The heat transfer  increases for nonuniform rotation with a positive Rossby number $ (\textrm{Ro}> 0) $.
\item With increasing the modulation frequency of $ \omega_{\textrm {mod}} $  the Nusselt number $ \Delta \textrm {Nu} $  decreases, which leads to suppression of heat transfer as with positive $ (\textrm{Ro}> 0) $, so with negative $ (\textrm{Ro} <0) $ rotation profiles.
\item The effect of increasing the modulation amplitude of $ \delta_{\textrm{mod}} $ is to increase the heat transfer  anyway of  the rotation profile.
\end{enumerate}
For the case of gravitational modulation, an increase  the Taylor number $ \textrm{T}_1 $ leads to a decrease in the variations of the Nusselt number $ \Delta \textrm{Nu} $ (Fig. \ref {fg6g} -Fig. \ref {fg6gr}). However, the heat flux increases (Fig. \ref {fg6r}-Fig. \ref {fg6rt})  for large numbers $ \textrm {T}_1 $ in the case of rotation modulation.  With increasing the Chandrasekhar number $ \textrm {Q}_1 $, heat transfer is at first suppresses and then increases (Fig. \ref {fg6m} - Fig. \ref{fg6mr}) for the case of modulation of the magnetic field.

Finally let us  compare the different types of parametric effects on a stationary   nonuniformly  rotating magnetoconvection among themselves.
In Fig. \ref {fg4}, we have depicted  the results of numerical solutions of the equations (\ref{eq43n}), (\ref{eq47n}), (\ref{eq50n}), (\ref{eq54n}) for fixed convection parameters: $ \textrm {Q} / \pi^2 = \textrm {Q}_1 = $ 80, $ \textrm {Ta} / \pi^4 = \textrm {T}_1 = 10^5 $, $ \textrm {Ra}_c / \pi^4 = \textrm {R}_1 = $ 9500, $ \textrm {Pm} = 1 $, $ \textrm {Pr} = 10 $, $ A_0 = 0.5 $. The frequencies and amplitudes of four types of modulation were considered equal: $ \omega_T = \omega_g = \omega_B = \omega_R = 10 $ and $ \delta_1 = \delta_2 = \delta_3 = \delta_4 = 0.3 $, phase $ \varphi = \pi $. In Fig. \ref {fg4}  we can observe that the rotational modulation has the greatest influence on the change in heat flow in the system  for positive $ (\textrm {Ro} = 2) $ and negative $ (\textrm {Ro} = - 3/4) $ rotation profiles. Gravity modulation slightly exceeds magnetic modulation: $ \Delta \textrm {Nu} |_ {\textrm {grav.mod}} \geq \Delta \textrm {Nu} |_ {\textrm {mag.mod}} $.
Thermal phase modulation has a lesser effect on heat transfer  in comparison with other types of modulations:
$$\Delta\textrm{Nu}|_{\textrm{temp.mod}}<\Delta\textrm{Nu}|_{\textrm{mag.mod}} \leq \Delta\textrm{Nu}|_{\textrm{grav.mod}}<\Delta\textrm{Nu}|_{\textrm{rot.mod}}. $$

\end{document}